\documentclass[aps,prl,showpacs,twocolumn,notitlepage,superscriptaddress,nofootinbib,longbibliography]{revtex4-2}
\usepackage{enumitem}

\usepackage{braket}
\usepackage{amssymb,amsfonts,amsmath,amsthm}
\usepackage[hidelinks]{hyperref}
\usepackage{booktabs}
\usepackage{multirow}
\usepackage{ulem}

\usepackage{mathtools}
  \mathtoolsset{showonlyrefs}

\newtheorem{theorem}{Theorem}
\newtheorem*{theorem*}{Theorem}
\newtheorem{proposition}[theorem]{Proposition}

\def\dwn{\downarrow}
\def\D{\mathcal{D}}
\def\L{\mathcal{L}}
\def\O{\mathcal{O}}
\def\R{\mathcal{R}}
\def\S{\mathcal{S}}
\def\U{\mathcal{U}}
\def\F{\mathcal{F}}

\def\eig{\operatorname{eig}}
\def\diag{\operatorname{diag}}

\def\supp{\operatorname{supp}}

\def\free{\rightarrow}

\def\pset{\Delta^\dwn_n}

\def\Real{\mathbb{R}}

\def\ts{\gamma^\mathrm{ts}}

\def\RI{\mathrm{RI}}

\def\vecx{\vec{x}}
\def\vecy{\vec{y}}
\def\vecz{\vec{z}}
\def\vect{\vec{t}}
\def\vecxp{\vec{x}^{\,\prime}}

\def\vecF{\vec{F}}
\def\vecG{\vec{G}}
\def\vectau{\vec{\tau}}

\def\ie{\textit{i.e.}, }
\def\argmin{\operatorname*{arg\,min}}

\newcommand{\widevectau}{\smash{\widehat{\vectau}}\vphantom{\vectau}}

\begin{document}

\title{Optimal Approximate Quantum Conversion from a Taut-String Principle in Majorization-Based Resource Theories}

\begin{abstract}
Optimal approximate state conversion is a fundamental task in quantum resource theories, yet existing solutions are largely tailored to specific distinguishability measures and resource theories. 
We show that every optimal approximate-conversion problem in the three paradigmatic majorization-based quantum resource theories of non-uniformity, bipartite pure-state entanglement, and pure-state coherence reduces exactly to one of only four elementary taut-string constructions. 
This holds for problems generated by Schur-convex spectral distances and by a broad family of quantum divergences satisfying a natural reduction property with respect to the corresponding equivalence group. 
The resulting framework encompasses a broad class of physically relevant distinguishability measures, including the trace distance, all unitarily invariant norms, quantum relative entropy, maximal quantum $f$-divergences, and the $\alpha$-$z$ Rényi relative entropies. 
The increment vector of the corresponding taut string completely determines both the optimal approximation error and a canonical optimal state. 
Our results establish a common geometric framework for approximate quantum resource conversion, revealing that a broad class of seemingly different optimization problems is governed by the same taut-string geometric principle.
\end{abstract}

\author{F. Meroi}
\affiliation{Instituto de Investigaciones Matemáticas Luis A. Santaló, Universidad de Buenos Aires, CONICET, Buenos Aires, Argentina}

\author{C. Massri}
\affiliation{Instituto de Investigaciones Matemáticas Luis A. Santaló, Universidad de Buenos Aires, CONICET, Buenos Aires, Argentina}

\author{M. Losada}
\affiliation{CONICET-Universidad de Buenos Aires, Instituto de Ciencias de la Computaci\'on (ICC), Buenos Aires, Argentina}

\author{G. Bellomo}
\affiliation{CONICET-Universidad de Buenos Aires, Instituto de Ciencias de la Computaci\'on (ICC), Buenos Aires, Argentina}

\author{G. M. Bosyk}
\email{gbosyk@icc.fcen.uba.ar (corresponding author)}
\affiliation{CONICET-Universidad de Buenos Aires, Instituto de Ciencias de la Computaci\'on (ICC), Buenos Aires, Argentina}

\date{\today}

	\maketitle

\par\indent\textit{Introduction}---
Quantum resource theories (QRTs) provide a unified framework for quantifying and manipulating physical resources under operational restrictions \cite{Chitambar2019,Gour2025}. A central task is determining whether a target state can be obtained from a given initial state using only the allowed free operations. Since exact state conversion is often impossible, the relevant operational problem becomes identifying the physically accessible state that best approximates the desired target according to a chosen distinguishability measure.

Optimal approximate conversion has therefore become a fundamental problem throughout quantum information science. It arises in entanglement manipulation, quantum coherence, thermodynamics, asymmetry, and numerous other resource theories, where a wide variety of optimization methods have been developed for different distinguishability measures and physical settings (see, e.g., \cite{Torgersen1989,Vidal2000,Li1989,Renes2016,Li2016,Chubb2018,Massri2020,Benac2023,Zanoni2024,Kondra2025,Theurer2025}). Despite this extensive literature, existing solutions remain largely measure dependent. Trace distance, Schatten norms, fidelity-based criteria, quantum relative entropies, and more general quantum divergences have all been treated through distinct analytical techniques, leaving open the question of whether these seemingly different optimization problems share a common geometric structure.

In this Letter, we answer this question affirmatively. We prove that every approximate-conversion problem arising in the three paradigmatic majorization-based quantum resource theories of non-uniformity, bipartite pure-state entanglement, and pure-state coherence belongs to one of only four elementary taut-string geometries. 
This applies to all problems generated by Schur-convex spectral distances and by quantum divergences admitting reduction to a strictly increasing function of a classical Csiszár $f$-divergence depending only on the corresponding representative probability vectors. 
Consequently, optimization problems generated by the trace distance, all unitarily invariant norms, quantum relative entropy, maximal quantum $f$-divergences, and the $\alpha$-$z$ Rényi relative entropies are all governed by the same geometric principle.

The key ingredient is the classical geometry of taut strings, originally introduced in constrained approximation and variational optimization \cite{Dantzig1971}. We show that the optimal approximating state is completely determined by the increment vector of an appropriate taut string. 
The distinguishability measure determines the parametrization of the taut-string problem, whereas the underlying feasible geometry is fixed by the majorization constraints.
The apparent diversity of approximate-conversion problems therefore reflects different representations of the same one-dimensional variational geometry rather than fundamentally different optimization principles.

Majorization-based quantum resource theories \cite{Bosyk2019} provide the ideal setting in which this geometric principle becomes exact. They include bipartite pure-state entanglement \cite{Nielsen1999}, pure-state coherence \cite{Du2015,Streltsov2017}, and non-uniformity \cite{Gour2015,Streltsov2018}, all of which share the same convertibility criterion based on majorization \cite{Bosyk2019}. This common order endows their conversion sets with a rich geometric structure that enables the corresponding approximation problems to be solved exactly through taut-string constructions.

\begin{figure}[t]
    \centering
    \includegraphics[width=.99\linewidth]{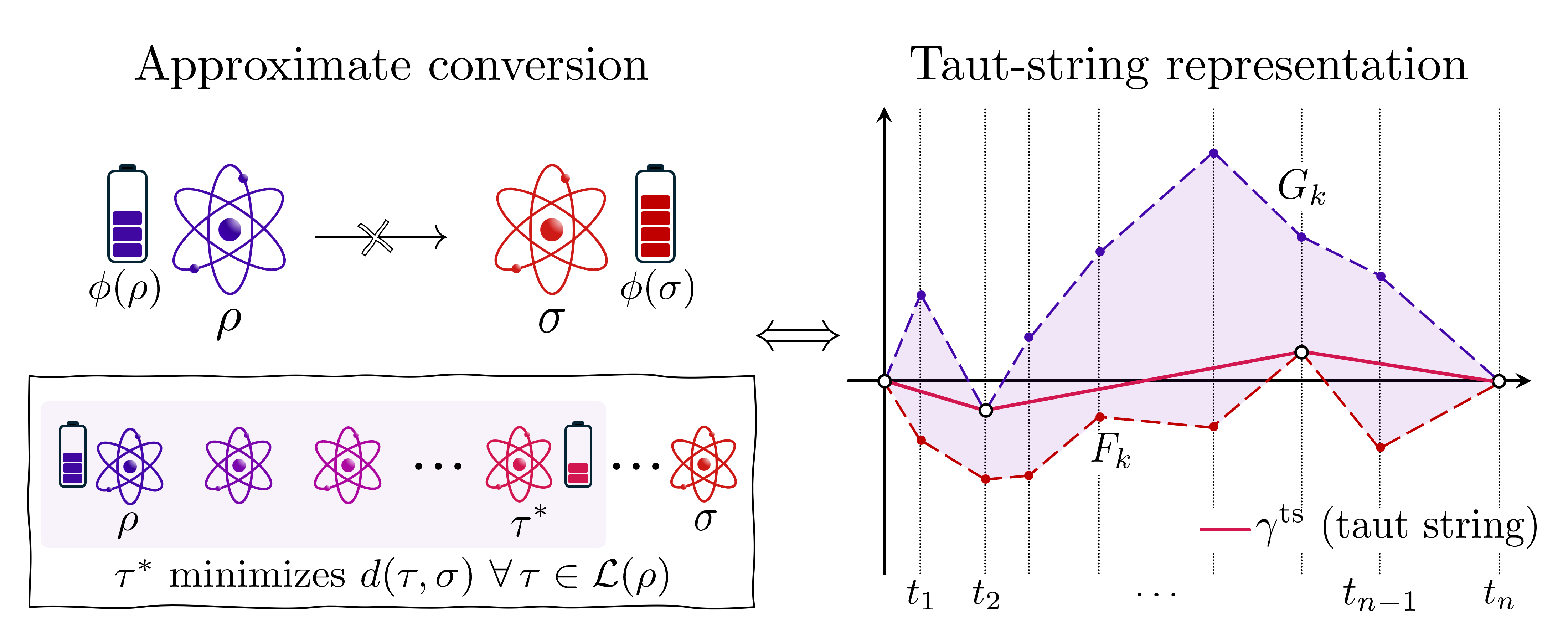}
    \caption{Schematic of optimal approximate conversion in MQRTs. The initial state ($\rho$) cannot be converted exactly into the target ($\sigma$) because it lies outside the reachable set under the resource preorder, represented here by the batteries. The best admissible approximation ($\tau$) is then selected by minimizing the chosen error measure $d$, encoded by the color gradient. In MQRTs, this problem is solved exactly by a taut-string construction constrained by the majorization floor ($F_k$) and roof ($G_k$). The construction for approximations over the upper conversion set is entirely analogous.}
    \label{fig:esquema}
\end{figure}

The overall reduction is illustrated in Fig.~\ref{fig:esquema}. The quantum approximate-conversion problem is first reduced to an optimization over the corresponding majorization polytope. We then show that every resulting optimization reduces to one of only four elementary taut-string geometries, from which both the optimal approximation error and the optimal approximating state are reconstructed directly.

Beyond yielding explicit solutions for a remarkably broad family of distinguishability measures, our results identify the geometric principle underlying approximate quantum resource conversion and establish taut-string methods as a new analytical tool for quantum information theory. This perspective unifies optimization problems previously regarded as unrelated and suggests that taut-string geometry may extend well beyond the majorization-based resource theories considered here. More generally, it opens new connections between majorization geometry, constrained variational optimization, and quantum information theory.

\par\indent\textit{Problem Formulation and Reduction to Majorization}---
Let $\S \subseteq \D(\mathbb{C}^{D})$ denote the relevant class of quantum states.
A quantum resource theory (QRT) is specified by a set of free states $\F\subset\S$ and a set of free quantum operations $\O$.
A deterministic conversion $\rho\free\sigma$ is possible whenever there exists $\Lambda\in\mathcal{O}$ such that $\sigma=\Lambda(\rho)$.

Whenever the desired transformation is impossible (\ie $\rho\not\free\sigma$), one seeks the admissible state that best approximates the target.
This naturally gives rise to the lower and upper conversion sets
\begin{equation}
\label{eq:conversion_set}
\L(\rho)=\{\tau\in\mathcal{S}:\rho\free\tau\},
\qquad
\U(\rho)=\{\tau\in\mathcal{S}:\tau\free\rho\},
\end{equation}
and to the corresponding optimization problems
\begin{equation}
\label{eq:problems}
\varepsilon^{\min}_{\L,d} = \inf_{\tau\in\L(\rho)} d(\tau,\sigma),
\qquad
\varepsilon^{\min}_{\U,d} = \inf_{\tau\in\U(\rho)} d(\tau,\sigma),
\end{equation}
where $d$ is a quantum distinguishability measure.
These two optimization problems play complementary operational roles: one searches for the closest state to $\sigma$ reachable from $\rho$, whereas the other searches for the closest state to $\sigma$ from which $\rho$ could have been obtained.

The present work focuses on majorization-based quantum resource theories (MQRTs).
An MQRT is characterized by the existence of a surjective representation map
\begin{equation}
    \R:\S\rightarrow\pset,
\end{equation}
for some $n$ (depending on the MQRT), such that deterministic state
conversion is completely characterized by majorization of the representative
probability vectors in the ordered probability simplex
\begin{equation}
    \pset=
    \left\{
    \vecx\in\Real^n:
    x_1\ge x_2\ge\cdots\ge x_n\ge0,\,
    \sum_{i=1}^n x_i=1
    \right\}.
\end{equation}
In particular, for non-uniformity and pure-state coherence, $n=D$. For bipartite
pure-state entanglement on $\mathcal H_A\otimes\mathcal H_B$,
$D=d_A d_B$ and $n=\min\{d_A,d_B\}$.

Furthermore, for the MQRTs considered here,
\begin{equation}
    \R(\rho)=\eig\circ\Phi(\rho),
\end{equation}
where $\eig(A)$ denotes the vector of eigenvalues of a Hermitian operator $A$, arranged in non-increasing order. 
Here $\Phi$ denotes the identity map for non-uniformity, the partial trace over the subsystem of larger dimension
for bipartite pure-state entanglement, and the dephasing map for pure-state coherence.

The majorization relation underlying this conversion criterion is conveniently expressed in terms of partial sums. 
For probability vectors $\vecx,\vecy\in\pset$, define the partial sums
\begin{equation}
s_k(\vecx)=\sum_{i=1}^{k}x_i, \qquad k=1,\ldots,n.
\end{equation}
We say that $\vecx$ majorizes $\vecy$, written $\vecx\succeq\vecy$, whenever
\begin{equation}
s_k(\vecx)\ge s_k(\vecy), \qquad k=1,\ldots,n-1,
\end{equation}
with equality for $k=n$ \cite{MarshallBook}.

Depending on the resource theory, the representation map is either order-preserving,
\begin{equation}
\rho\free\sigma \iff \R(\rho)\succeq\R(\sigma),
\end{equation}
or order-reversing,
\begin{equation}
\rho\free\sigma \iff \R(\sigma)\succeq\R(\rho).
\end{equation}
The former corresponds to non-uniformity \cite{Gour2015,Streltsov2018}, whereas pure-state entanglement \cite{Nielsen1999} and pure-state coherence \cite{Du2015,Zhu2017} belong to the latter class.

Introducing the lower and upper majorization polytopes \cite{Junior2022},
\begin{equation}
\label{eq:majorization_polytopes}
\L(\vecx)=\{\vecz\in\pset:\vecx\succeq\vecz\}, \qquad \U(\vecx)=\{\vecz\in\pset:\vecz\succeq\vecx\},
\end{equation}
the quantum optimization problems in Eq.~\eqref{eq:problems} reduce to optimization over one of these two polytopes, depending on whether the representative map is order-preserving or order-reversing.
The problem is therefore completely determined once the corresponding classical objective function is identified.

The central structural assumption of this work is that the quantum distinguishability measure reduces, after optimization over the reversible free transformations, to a classical distinguishability measure depending only on the representative probability vectors.
More precisely, let $G\subseteq U(D)$ denote the unitary group implementing reversible free transformations associated with the corresponding MQRT. 
We assume that
\begin{equation}
\label{eq:reduction_measures_property}
\inf_{U\in G} d(U\rho U^\dagger,\sigma) = C(\vecx,\vecy),
\end{equation}
where $\vecx=\R(\rho)$, $\vecy=\R(\sigma)$, and $C$ is a classical distinguishability measure.

The error measures considered in this Letter all satisfy Eq.~\eqref{eq:reduction_measures_property}.
They fall into two broad classes.

The first consists of spectral distances of the form
\begin{equation}
\label{eq:quantum_Schur_convex_dist}
d_g(\rho,\sigma) = g\left(\eig(\sigma-\rho)\right),
\end{equation}
where $g:\Real^D\rightarrow\Real$ is Schur-convex and such that $d_g$ defines a distance.
This class includes all distances induced by unitarily invariant norms, including the Schatten norms and, in particular, the trace distance.

The second class, which we use for non-uniformity, consists of quantum divergence measures $S$ whose optimization over unitary orbits reduces to a classical divergence,
\begin{equation}
\label{eq:quantum_divergence_reduction}
\inf_{U\in U(D)} S\left(U\rho U^\dagger \middle\| \sigma \right) = h\left(D_f(\vecx\|\vecy)\right),
\end{equation}
where
\begin{equation}
\label{eq:classical_f_divergence}
D_f(\vecx\|\vecy) = \sum_{i=1}^{n} y_i f\left(\frac{x_i}{y_i}\right)
\end{equation}
is the classical Csiszár $f$-divergence \cite{Csiszar1967}, $f:\Real_+\rightarrow\Real$ is convex with the convention $0f\left(x/0\right) = x\lim_{t\rightarrow\infty}\frac{f(t)}{t}$ for $x>0$ and $0f\left(0/0\right) = 0$, and $h:\Real\cup\{+\infty\}\rightarrow\Real\cup\{+\infty\}$ is strictly increasing with $h(+\infty)=+\infty$.
This reduction property is shared by a broad class of quantum divergences, including the quantum relative entropy \cite{Zhang2014}, maximal quantum $f$-divergences \cite{Matsumoto2024,Nguyen2026}, the $\alpha$-$z$ Rényi relative entropies \cite{Audenaert2015,Trinh2025}, and several other quantum divergence families \cite{Li2016,Dinh2024,Li2025}.

Consequently, only two classical objective functions remain.

For spectral distances,
\begin{equation}
\label{eq:classical_g_non-unif}
C(\vecx,\vecy) = g(\vecy-\vecx)
\end{equation}
for non-uniformity, whereas for pure-state entanglement and coherence,
\begin{equation}
\label{eq:classical_g_ent_coh}
C(\vecx,\vecy) = g\left(D_{\RI}(\vecx,\vecy), -D_{\RI}(\vecx,\vecy),\vec{0}_{D-2} \right),
\end{equation}
where $D_{RI}$ denotes the root-infidelity distance,
\begin{equation}
\label{eq:classical_Schur_convex_dist_ent_coh}
D_{\RI}(\vecx,\vecy) = \sqrt{1-F^2(\vecx,\vecy)}, \qquad F(\vecx,\vecy)=\sum_i\sqrt{x_i y_i}.
\end{equation}

For quantum divergences, which provide an additional class only for non-uniformity\footnote{For pure-state entanglement and coherence, the quantum divergences considered here either diverge or reduce to monotone functions of the fidelity $F$, thereby recovering the spectral-distance setting.},
the classical objective function is
\begin{equation}
\label{eq:classical_div}
C(\vecx,\vecy) = h\left(D_f(\vecx\|\vecy)\right).
\end{equation}

\par\indent\textit{Taut-string characterization}---
Although different resource theories, conversion directions, and distinguishability measures lead \textit{a priori} to different optimization problems, every one of them falls into one of only four elementary taut-string geometries.
Consequently, the optimal approximation error and an optimal approximating state are completely determined by the increment vector of a one-dimensional taut string.

\begin{figure}[t]
\centering
\includegraphics[width=.98\linewidth]{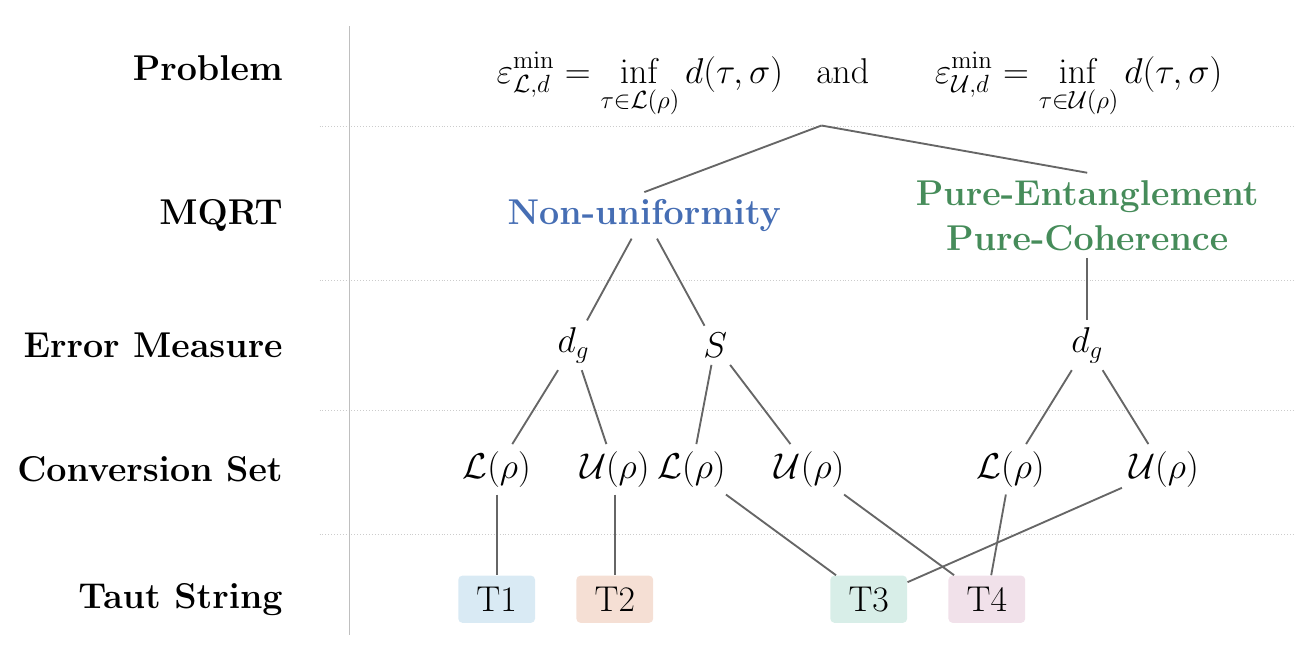}
\caption{
Classification of optimal approximate-conversion problems by taut-string geometry.
Every optimization problem considered in this Letter reduces to one of only four elementary taut-string constructions.
The increment vector of the corresponding taut string determines both the optimal approximation error and an optimal approximating state.
}
\label{fig:esquema_thm}
\end{figure}

Figure~\ref{fig:esquema_thm} summarizes this classification.
The resource theory and conversion direction determine the corresponding majorization polytope, while the distinguishability measure specifies the parametrization of the taut-string problem.
Once the appropriate taut string has been constructed, its increment vector completely determines the solution of the original quantum optimization problem.

The underlying geometric object is the classical taut string \cite{Dantzig1971}.
Given a strictly increasing sequence $\vect=(t_0,\ldots,t_n)$ and boundary vectors $\vecF$ (Floor) and $\vecG$ (Roof) in $\Real^{n+1}$ satisfying $F_k\leq G_k$ for $k=1,...,n-1$, $F_0=G_0$, and $F_n=G_n$, let $\Gamma_{\vecF,\vecG}$ denote the set of continuous piecewise-linear curves $\gamma:[t_0,t_n]\rightarrow\Real$ whose nodal values satisfy
$F_k\le\gamma(t_k)\le G_k$ for $k=0,\ldots,n$.

The taut string is the unique minimizer of the Euclidean length functional
\begin{equation}
\ts = \argmin_{\gamma\in\Gamma_{\vecF,\vecG}(\vect)} \sum_{k=1}^{n}  \sqrt{(t_k-t_{k-1})^2+ \left(\gamma(t_k)-\gamma(t_{k-1})\right)^2},
\end{equation}
and its increment vector $\vectau=(\tau_1, \ldots, \tau_n)$ is given by
\begin{equation}
\tau_k = \ts(t_k)-\ts(t_{k-1}), \qquad k=1,\ldots,n.
\end{equation}

Although originating in constrained approximation theory, taut strings admit efficient computational implementations \cite{Davies2001}.

We can now state the main result, which shows that optimal approximate quantum conversion is governed by a common taut-string geometric principle.

\begin{theorem*}[Taut-string characterization of optimal approximate-conversion problems]
\label{thm:main_theorem}
Let the resource theory be bipartite pure-state entanglement, pure-state coherence, or non-uniformity, and let $\rho,\sigma\in\S$ with $\rho$ and $\sigma$ pure in the entanglement and coherence cases and $\rho\not\free\sigma$.
Write $\vecx=\R(\rho)$ and $\vecy=\R(\sigma)$ for their representative probability vectors.

Then every approximate-conversion problem introduced in Eq.~\eqref{eq:problems}, generated by the distinguishability measures considered in this Letter, spectral distances $d_g$ given in \eqref{eq:quantum_Schur_convex_dist} and quantum divergences $S$ satisfying \eqref{eq:quantum_divergence_reduction}, reduces exactly to one of the four elementary taut-string geometries listed in Table~\ref{tab:taut_string_cases}.

In every case, the increment vector of the corresponding taut string completely determines both the optimal approximation error and an optimal approximating state.

\end{theorem*}

\begin{proof}
The proof proceeds in two steps.

First, Proposition~\ref{prop:reduction_majorization_polytopes} reduces every quantum approximate-conversion problem considered here to an optimization over a lower or upper
majorization polytope.

For non-uniformity, the representation map is order preserving, so that
\begin{equation}
    \epsilon^{\min}_{\L,d} =  \inf_{\vecx'\in \L(\vecx)} C(\vecx',\vecy),
    \qquad
    \epsilon^{\min}_{\U,d} = \inf_{\vecx'\in \U(\vecx)} C(\vecx',\vecy).
\end{equation}
For spectral distances,
\begin{equation}
    C(\vecx,\vecy)=g(\vecy-\vecx),
\end{equation}
where in the non-uniformity case $n=D$.
Therefore, item~\ref{item:prop2_schur} of  Prop.~\ref{prop:reduction_taut_string} with $\varphi=g$ shows that the lower- and upper-set problems are solved by geometries T1 and T2, respectively.

For quantum divergences,
\begin{equation}
    C(\vecx,\vecy)   =   h\!\left(D_f(\vecx\Vert\vecy)\right),
\end{equation}
with $h$ strictly increasing.
Hence minimizing $C$ is equivalent to minimizing $D_f$, and item~\ref{item:prop2_fdiv} of Prop.~\ref{prop:reduction_taut_string} shows that the lower- and upper-set problems are solved by geometries T3 and T4, respectively.

We now consider bipartite pure-state entanglement and pure-state coherence. 
In these MQRTs the representation map is order reversing, and Proposition~\ref{prop:reduction_majorization_polytopes} gives
\begin{equation}
    \epsilon^{\min}_{\L,d_g}   =  \inf_{\vecx'\in \U(\vecx)}   C(\vecx',\vecy),
    \qquad
    \epsilon^{\min}_{\U,d_g}   =  \inf_{\vecx'\in L(\vecx)}   C(\vecx',\vecy),
\end{equation}
where
\begin{equation}
    C(\vecx,\vecy)   = g\!\left(D_{\RI}(\vecx,\vecy), -D_{\RI}(\vecx,\vecy),\vec{0}_{D-2} \right).
\end{equation}

To identify the minimizers of this objective, consider the convex generator
\begin{equation}
    f_H(u)=(\sqrt{u}-1)^2.
\end{equation}
Its associated Csisz\'ar $f$-divergence satisfies
\begin{equation}
    D_{f_H}(\vecx'\Vert\vecy)   =  2\left[1-F(\vecx',\vecy)\right].
\end{equation}
Consequently,
\begin{equation}
    D_{\rm RI}(\vecx',\vecy)   =  r\!\left(D_{f_H}(\vecx'\Vert\vecy)\right),
    \qquad
    r(u)=\sqrt{u-\frac{u^2}{4}}.
\end{equation}
Since $r$ is non-decreasing on $[0,2]$, every minimizer of $D_{f_H}(\vecx'\Vert\vecy)$ is also a minimizer of $D_{\RI}(\vecx',\vecy)$.

Moreover, defining
\begin{equation}
    q_g(t)  = g\!\left(t,-t,\vec{0}_{D-2}\right), \qquad t\ge0,
\end{equation}
Schur-convexity of $g$ implies that $q_g$ is non-decreasing, since for $0\leq s\leq t$,
\begin{equation}
    (t,-t,\vec{0}_{D-2})  \succeq    (s,-s,\vec{0}_{D-2}).
\end{equation}
Therefore, every minimizer of $D_{\RI}$ also minimizes $C(\vecx',\vecy)=q_g(D_{\rm RI}(\vecx',\vecy))$.

Applying item~\ref{item:prop2_fdiv} of Prop.~\ref{prop:reduction_taut_string} to $f=f_H$, geometry T3 provides a
minimizer over $\L(\vecx)$, namely $\widevectau^{(3)}$, whereas geometry T4 provides a minimizer over $\U(\vecx)$, namely $\widevectau^{(4)}$.

Finally, because the representation map is order reversing for pure-state entanglement and coherence, the classical upper-set
solution T4 corresponds to the quantum optimization over $\L(\rho)$, whereas the classical lower-set solution T3 corresponds to the
quantum optimization over $\U(\rho)$.

Thus every approximate-conversion problem considered in the theorem is solved by exactly one of the four taut-string geometries T1--T4,
and the corresponding increment vector determines both the optimal error and a canonical optimal approximating state.
\end{proof}

\begin{table}[b]
\begin{tabular}{cccc}
\toprule
Geometry & $t_k$ &  $F_k$ (Floor)&  $G_k$ (Roof)\\
\midrule
T1 & $k/n$ & $s_k(\vecy-\vecx)$ & $1$ ($0$ at $k=0,n$)\\
T2 & $k/n$ & $s_k(\vecy)-1$ ($k\ge1$, $0$ at $k=0$) & $s_k(\vecy-\vecx)$\\
T3 & $s_k(\vecy)$ & $0$ ($s_m(\vecx)$ at $k=m$)& $s_k(\vecx)$\\
T4 & $s_k(\vecy)$ & $s_k(\vecx)$  ($1$ at $k=m$)& $1$ ($0$ at $k=0$)\\
\bottomrule
\end{tabular}
\caption{The four elementary taut-string geometries appearing in the Theorem, from which the corresponding increment vectors $\vectau^{(j)}$ are derived.
For geometries T1 and T2, $k$ ranges from $0$ to $n$. 
For geometries T3 and T4, $k$ ranges from $0$ to $m$ with $m=|\supp(\vecy)| \leq n$, so that, when $m<n$, we define the increment vector extensions $\widevectau^{(3)} = \vectau^{(3)} \oplus (x_{m+1},\ldots,x_n)$ and $\widevectau^{(4)} = \vectau^{(4)} \oplus \vec{0}_{n-m}$, for geometries T3 and T4, respectively.
See the Supplemental Material for further details \cite{suppMat}.
}
\label{tab:taut_string_cases}
\end{table}

The theorem immediately yields explicit solutions for the three MQRTs considered in this work.

Given an approximate-conversion problem, the resource theory, conversion direction, and distinguishability measure uniquely determine which of the four geometries T1--T4 applies.
Once the corresponding taut string has been constructed, its increment vector directly specifies both the optimal approximation error and the optimal approximating state in canonical form (Table~\ref{tab:optimal_states}).

\smallskip
\noindent
\textit{Non-uniformity.}
For Schur-convex spectral distances,
\begin{equation}
\varepsilon^{\min}_{X,d_g} = g(\vectau^{(j)}),
\end{equation}
where $\vectau^{(1)}$ is obtained from geometry T1 for $X=\L(\rho)$ and $\vectau^{(2)}$ from geometry T2 for $X=\U(\rho)$.

For quantum divergences,
\begin{equation}
\varepsilon^{\min}_{X,S} = h\left(D_f(\widevectau^{(j)}\|\vecy)\right),
\end{equation}
where $\vectau^{(j)}$ is obtained from geometry T3 for $X=\L(\rho)$ and from geometry T4 for $X=\U(\rho)$.

\smallskip
\noindent
\textit{Entanglement and coherence.}
For spectral distances,
\begin{equation}
\varepsilon^{\min}_{X,d_g} = q_g\left(D_{\RI}(\widevectau^{(j)},\vecy)\right) \ \text{with } q_g(t)=g(t,-t,\vec{0}_{D-2}),
\end{equation}
where $\widevectau^{(3)}= \vectau^{(3)} \oplus (x_{m+1},\ldots,x_n)$ is obtained from geometry T3 for $X=\U(\rho)$ and $\widevectau^{(4)}=\vectau^{(4)} \oplus \vec{0}_{n-m}$ from T4 for $X=\L(\rho)$.

For all three MQRTs, the increment vector $\vectau$ also determines the optimal approximating state. 
We express this state in the basis naturally associated with the target state: the eigenbasis for non-uniformity, the Schmidt basis for bipartite entanglement, and the incoherent basis for coherence. 
The resulting canonical forms are summarized in Table~\ref{tab:optimal_states}.

\begin{table}[b]
    \centering
    \small
    \setlength{\tabcolsep}{3pt}
    \renewcommand{\arraystretch}{1.15}

    \begin{tabular}{@{}lccc@{}}
        \toprule
        MQRT
        & $d$
        & Optimal error $\varepsilon^{\min}$
        & Optimal state
        \\
        \midrule

        \multirow{2}{*}{Non-uniformity}
        & $d_g$
        & $g\!\left(\vectau^{(j)}\right)$
        & $\diag\!\left(\vecy-\vectau^{(j)}\right)$
        \\

        & $S$
        & $h\!\left(D_f(\widevectau^{(j)}\|\vecy)\right)$
        & $\diag\!\left(\widevectau^{(j)}\right)$
        \\

        \midrule

        \begin{tabular}[c]{@{}l@{}}
            Entanglement\\
            Coherence
        \end{tabular}
        &
        $d_g$
        &
        $q_g\!\left(
            D_{\RI}(\widevectau^{(j)},\vecy)
        \right)$
        &
        $\begin{gathered}
            \displaystyle
            \sum_i \sqrt{\widevectau^{(j)}_i}\ket{ii}
            \\[1mm]
            \displaystyle
            \sum_i \sqrt{\widevectau^{(j)}_i}\ket{i}
        \end{gathered}$
        \\

        \bottomrule
    \end{tabular}

    \caption{
        Optimal approximation errors and canonical optimal states determined by the taut-string increment vectors
        $\vectau^{(j)}$ and, when required, their extensions $\widevectau^{(j)}$, with $j$ determined by the MQRT, the distinguishability measure and
        conversion set as indicated in Fig.~\ref{fig:esquema_thm}.
        Here, $q_g(t)=g\!\left(t,-t,\vec{0}_{D-2}\right)$.
        For non-uniformity, the canonical form is expressed in the eigenbasis of the target state $\sigma$; for bipartite
        pure-state entanglement, in its Schmidt basis; and for pure-state coherence, in the incoherent basis, with the
        target assumed to be in canonical form with non-negative amplitudes.
    }
    \label{tab:optimal_states}
\end{table}

\par\indent\textit{Discussion and Outlook}---
The results of this Letter reveal the geometric principle underlying optimal approximate quantum resource conversion.
Optimization problems that have traditionally been studied separately for different distinguishability measures and different quantum resource theories are shown to belong to the same geometric class.
For the three paradigmatic majorization-based quantum resource theories---non-uniformity, bipartite pure-state entanglement, and pure-state coherence---every approximate-conversion problem considered here reduces to one of only four elementary one-dimensional taut-string constructions.
Consequently, both the optimal approximation error and the optimal approximating state are completely determined by the increment vector of the corresponding taut string.

This changes the usual perspective on approximate quantum resource conversion.
Rather than requiring a different optimization method for each distinguishability measure, the resource preorder fixes the feasible geometry through the associated majorization polytope, while the distinguishability measure determines the corresponding taut-string parametrization and the classical objective function.
Trace distance, all unitarily invariant norms, quantum relative entropy, maximal quantum $f$-divergences, and the $\alpha$-$z$ Rényi relative entropies therefore become different manifestations of the same underlying variational structure.

Beyond providing explicit analytical solutions for non-uniformity, bipartite pure-state entanglement, and pure-state coherence, our results establish taut-string geometry as a new analytical tool for constrained optimization in quantum information theory.
A natural next step is to investigate whether analogous constructions exist for resource theories governed by generalized majorization relations, including thermo-majorization \cite{Ruch1978,Horodecki2013} and catalytic transformations \cite{Klimesh2007,Turgut2007,Datta2023,Bartosik2024,Elkouss2026}, where cumulative-order constraints also arise naturally.

More broadly, the emergence of taut strings in quantum resource conversion suggests that constrained shortest-path geometry may provide a general mathematical framework for optimization under majorization-type constraints.
Since such constraints appear throughout statistical physics, information theory, economics, and applied mathematics \cite{MarshallBook,Arnold2007}, the geometric framework developed here may find applications well beyond quantum resource theories.

\begin{acknowledgments}
This work is partially supported by Consejo Nacional de Investigaciones Cient\'ficas y T\'ecnicas CONICET (Argentina) through project PIP 2023-2025 Nro. 0135, and Universidad de Buenos Aires UBA (Argentina) through project UBACyT Programación Científica 2026 and by the European Union through the MSCA SE project QCOMICAL (Grant Agreement ID:101182520).
\end{acknowledgments}

\appendix

\section{Appendix: the two reductions}

The proof of the main theorem proceeds in two independent steps: 
Proposition~\ref{prop:reduction_majorization_polytopes} reduces every quantum approximate-conversion problem considered in this Letter to the optimization of an appropriate classical objective function over a majorization polytope.
Proposition~\ref{prop:reduction_taut_string}, which is of mathematical interest in itself, solves the two classical optimization primitives needed in the proof: Schur-convex difference objectives and Csiszár $f$-divergence objectives. 
Combined with the monotonic reductions established in the proof of the theorem, these two propositions yield the four taut-string geometries.

\begin{proposition}[Reduction to classical optimization over majorization polytopes]
\label{prop:reduction_majorization_polytopes}

Let the resource theory be bipartite pure-state entanglement, pure-state coherence, or non-uniformity, and let $\rho,\sigma\in\S$ with $\rho$ and $\sigma$ pure in the entanglement and coherence cases and $\rho\not\free\sigma$.
Write $\vecx=\R(\rho)$ and $\vecy=\R(\sigma)$ for their representative probability vectors.

Then every approximate-conversion problem introduced in Eq.~\eqref{eq:problems}, generated by the distinguishability measures considered in this Letter, reduces exactly to the optimization of the corresponding classical objective function $C(\vecx,\vecy)$  over the appropriate majorization polytope $\L(\vecx)$ or $\U(\vecx)$.

For non-uniformity,
\begin{equation}
\varepsilon^{\min}_{\L,d} = \inf_{\vecxp\in\L(\vecx)} C(\vecxp,\vecy),
\qquad
\varepsilon^{\min}_{\U,d} = \inf_{\vecxp\in\U(\vecx)} C(\vecxp,\vecy),
\end{equation}
where
\begin{equation}
C(\vecx,\vecy)=
\begin{cases}
g(\vecy-\vecx), & d=d_g,\\[1mm]
h\left(D_f(\vecx\|\vecy)\right), & d=S.
\end{cases}
\end{equation}

For pure-state entanglement and coherence,
\begin{equation}
\varepsilon^{\min}_{\L,d_g} = \inf_{\vecxp\in\U(\vecx)} C(\vecxp,\vecy),
\qquad
\varepsilon^{\min}_{\U,d_g} = \inf_{\vecxp\in\L(\vecx)} C(\vecxp,\vecy),
\end{equation}
where
\begin{equation}
C(\vecx,\vecy) = g\left(D_{\RI}(\vecx,\vecy),-D_{\RI}(\vecx,\vecy),\vec{0}_{D-2}\right).
\end{equation}

\end{proposition}

Hence every quantum approximate-conversion problem considered here reduces to a classical optimization problem over a majorization polytope. 
The following proposition identifies the taut-string solutions of the classical optimization primitives required in the proof of the theorem.

\begin{proposition}[Reduction to taut-string geometries]
\label{prop:reduction_taut_string}
Let $\vecx,\vecy\in\pset$ with  $m=|\supp(\vecy)| \leq n$.

\begin{enumerate}[label=\textup{(\roman*)}, leftmargin=*]

\item \label{item:prop2_schur} Schur-convex objectives. 

For every Schur-convex function $\varphi:\Real^n\rightarrow\Real$,
\begin{equation}
    \min_{\vecx'\in \L(\vecx)}   \varphi(\vecy-\vecx')    =    \varphi\!\left(\vec{\tau}^{(1)}\right),
\end{equation}
where $\vec{\tau}^{(1)}$ is the increment vector associated with geometry T1, and 
\begin{equation}
    \vecy-\vec{\tau}^{(1)}\in \L(\vecx)
\end{equation}
is a minimizer.

Similarly,
\begin{equation}
    \min_{\vecx'\in \U(\vecx)}    \varphi(\vecy-\vecx')     =    \varphi\!\left(\vec{\tau}^{(2)}\right),
\end{equation}
where $\vec{\tau}^{(2)}$ is the increment vector associated with geometry T2, and
\begin{equation}
    \vecy-\vec{\tau}^{(2)}\in \U(\vecx)
\end{equation}
is a minimizer.

\item \label{item:prop2_fdiv} $f$-divergence objectives.

Let $f:\mathbb{R}_{+}\rightarrow\mathbb{R}$ be convex, and let $D_f$ be the corresponding Csisz\'ar $f$-divergence.
Then
\begin{equation}
    \min_{\vecx'\in \L(\vecx)}
    D_f(\vecx'\Vert\vecy)    =    D_f\!\left(\widevectau^{(3)}  \Vert \vecy  \right),
\end{equation}
where $\vec{\tau}^{(3)}$ is the increment vector associated with geometry T3, and
\begin{equation}
    \widevectau^{(3)}    =   \vec{\tau}^{(3)} \oplus  (x_{m+1},\ldots,x_n) \in \L(\vecx)
\end{equation}
is a minimizer.

Similarly,
\begin{equation}
    \min_{\vecx'\in \U(\vecx)}    D_f(\vecx'\Vert\vecy) =    D_f\!\left(\widehat{\vec{\tau}}^{(4)}  \Vert\vecy   \right),
\end{equation}
where $\vec{\tau}^{(4)}$ is the increment vector associated with geometry T4, and
\begin{equation}
    \widevectau^{(4)}    =   \vec{\tau}^{(4)} \oplus \vec{0}_{n-m}  \in \U(\vecx)
\end{equation}
is a minimizer.
When $m=n$, the extensions are trivial.

\end{enumerate}

\end{proposition}

The proofs of Propositions~\ref{prop:reduction_majorization_polytopes} and~\ref{prop:reduction_taut_string}, and an illustrative example are provided in the Supplemental Material~\cite{suppMat}.

\bibliography{references}


\end{document}


\title{Supplementary Material}

\date{\today}
\maketitle

\tableofcontents

\section{Notation, Conventions, and Roadmap}
\label{sec:notation_roadmap}

We first collect the notation and conventions used throughout the Supplemental Material.
For a vector $\vecu\in\Real^n$, we denote by $\vecu^\dwn$ the vector obtained by arranging its components in non-increasing order, and define the partial sums
\begin{equation}
s_k(\vecu)=\sum_{i=1}^{k}u_i,
\qquad
k=1,\ldots,n,
\end{equation}
with the convention $s_0(\vecu)=0$.

Although the representative vectors of the MQRTs considered in the Letter belong to the ordered probability simplex
\begin{equation}
\pset
=
\left\{
\vecx\in\Real^n:
x_1\geq\cdots\geq x_n\geq0,
\quad
s_n(\vecx)=1
\right\},
\end{equation}
we will also apply majorization to real vectors that need not be probability vectors.
For $\vecu,\vecv\in\Real^n$ with $s_n(\vecu)=s_n(\vecv)$, we write
$\vecu\succeq\vecv$ whenever
\begin{equation}
s_k(\vecu^\dwn)\geq s_k(\vecv^\dwn),
\qquad
k=1,\ldots,n-1.
\end{equation}
For vectors in $\pset$, this reduces to the usual majorization relation used in the Letter.

For $\vecx\in\pset$, the lower and upper majorization polytopes are
\begin{equation}
\L(\vecx)
=
\left\{
\vecz\in\pset:\vecx\succeq\vecz
\right\},
\qquad
\U(\vecx)
=
\left\{
\vecz\in\pset:\vecz\succeq\vecx
\right\}.
\end{equation}
Throughout the Supplemental Material, $\vecx=\R(\rho)$ and
$\vecy=\R(\sigma)$ denote the representative probability vectors of the
initial and target states, respectively.

For the pure-state entanglement and coherence problems, we use the classical fidelity
\begin{equation}
F(\vecx,\vecy)
=
\sum_{i=1}^{n}\sqrt{x_i y_i},
\end{equation}
and the corresponding root-infidelity distance
\begin{equation}
D_{\RI}(\vecx,\vecy)
=
\sqrt{1-F^2(\vecx,\vecy)}.
\end{equation}

For the $f$-divergence constructions, let
\begin{equation}
m=|\supp(\vecy)|.
\end{equation}
Since $\vecy\in\pset$, this means that $y_k>0$ for $k=1,\ldots,m$ and
$y_k=0$ for $k=m+1,\ldots,n$. Consequently, the nodes
$t_k=s_k(\vecy)$, $k=0,\ldots,m$, are strictly increasing.

The taut-string geometries T3 and T4 therefore produce increment vectors
with $m$ components when $m<n$. To distinguish these geometric increment
vectors from the $n$-dimensional probability vectors entering the original
optimization problems, we use the extensions
\begin{align}
\widevectau^{(3)}
&=
\vectau^{(3)}
\oplus
(x_{m+1},\ldots,x_n),
\label{eq:T3_extension}
\\
\widevectau^{(4)}
&=
\vectau^{(4)}
\oplus
\vec{0}_{n-m}.
\label{eq:T4_extension}
\end{align}
When $m=n$, these extensions are trivial. In what follows,
$\vectau^{(j)}$ denotes the increment vector of geometry T$j$, while
$\widevectau^{(j)}$ denotes its $n$-dimensional extension whenever
such an extension is required. When no ambiguity can arise, the geometry
label will be omitted.

The four taut-string constructions appearing in the Letter solve the four classical optimization primitives summarized in Table~\ref{tab:roadmap}.
For the Schur-convex objectives, the optimal representative vector is obtained by subtracting the taut-string increment vector from the target vector. 
Section~\ref{sec:schur_taut} establishes the corresponding result in the more general setting of an arbitrary Schur-convex objective function.
For the $f$-divergence objectives, the extended increment vector itself is the optimal representative vector.

\begin{table}[t]
\centering
\begin{tabular}{c c c c}
\toprule
Geometry
&
Classical optimization problem
&
Nodes $t_k$
&
Optimal vector
\\
\midrule
T1
&
$\displaystyle
\min_{\vecxp\in\L(\vecx)}
g(\vecy-\vecxp)$
&
$k/n$
&
$\vecy-\vectau^{(1)}$
\\[3mm]
T2
&
$\displaystyle
\min_{\vecxp\in\U(\vecx)}
g(\vecy-\vecxp)$
&
$k/n$
&
$\vecy-\vectau^{(2)}$
\\[3mm]
T3
&
$\displaystyle
\min_{\vecxp\in\L(\vecx)}
D_f(\vecxp\|\vecy)$
&
$s_k(\vecy)$
&
$\widevectau^{(3)}$
\\[3mm]
T4
&
$\displaystyle
\min_{\vecxp\in\U(\vecx)}
D_f(\vecxp\|\vecy)$
&
$s_k(\vecy)$
&
$\widevectau^{(4)}$
\\
\bottomrule
\end{tabular}
\caption{
Roadmap of the four elementary taut-string geometries.
For T1 and T2, $k=0,\ldots,n$, whereas for T3 and T4,
$k=0,\ldots,m$, with $m=|\supp(\vecy)|$.
The precise floor and roof boundaries defining each geometry are derived
in the corresponding sections below.
}
\label{tab:roadmap}
\end{table}

The proof of the main result proceeds in two steps, corresponding to Propositions~1 and~2 of the Letter. 
Section~\ref{sec:prop1} proves Proposition~1, reducing the quantum approximate-conversion problems to classical optimization over the appropriate lower or upper majorization polytope. 
Section~\ref{sec:taut_preliminaries} collects the general taut-string definitions and properties used throughout the subsequent analysis. 
Section~\ref{sec:schur_taut} derives geometries T1 and T2 for Schur-convex objectives. 
Section~\ref{sec:d-maj} introduces the required $\vec{d}$-majorization, $f$-divergence, and Lorenz-curve machinery, and Sec.~\ref{sec:fdiv_taut} derives geometries T3 and T4 for Csisz\'ar
$f$-divergences. 
Section~\ref{sec:prop2} combines these results to prove Proposition~2. 
Together with the monotone fidelity reduction used in the proof of the main theorem in the Letter, these results yield the four taut-string solutions of the quantum approximate-conversion
problems. 
Finally, Sec.~\ref{sec:example} illustrates the four constructions with a numerical example.

\section{Proof of Proposition 1: Reduction to Classical Optimization over Majorization Polytopes}
\label{sec:prop1}

\begin{proposition}[Reduction to classical optimization over majorization polytopes]
\label{prop:prop1}
Let the resource theory be bipartite pure-state entanglement, pure-state coherence, or non-uniformity, and let $\rho,\sigma\in\S$, with $\rho$ and $\sigma$ pure in the entanglement and coherence cases and $\rho\not\free\sigma$.
Write $\vecx=\R(\rho)$ and $\vecy=\R(\sigma)$ for their representative probability vectors.

Consider the approximate-conversion problems
\begin{equation}
\label{eq:problems}
\varepsilon^{\min}_{\L,d}
=
\inf_{\tau\in\L(\rho)} d(\tau,\sigma),
\qquad
\varepsilon^{\min}_{\U,d}
=
\inf_{\tau\in\U(\rho)} d(\tau,\sigma).
\end{equation}
For all three resource theories, let $d=d_g$ be a spectral distance of the form
\begin{equation}
\label{eq:quantum_Schur_convex_dist}
d_g(\rho,\sigma)
=
g\left(\eig(\sigma-\rho)\right),
\end{equation}
where $g:\Real^D\rightarrow\Real$ is Schur-convex and $d_g$ defines a distance.
In addition, for non-uniformity, let $d=S$ be a quantum divergence satisfying
\begin{equation}
\label{eq:quantum_divergence_reduction}
\inf_{U\in U(D)}
S\left(U\rho U^\dagger\middle\|\sigma\right)
=
h\left(D_f(\vecx\|\vecy)\right),
\end{equation}
where $D_f$ is a classical Csiszár $f$-divergence generated by a convex function $f$, and $h:\Real \cup \{+\infty\}\rightarrow\Real\cup \{+\infty\}$ is strictly increasing with $h(+\infty)=+\infty$.

Then each problem in Eq.~\eqref{eq:problems} reduces exactly to the optimization of a classical objective function $C$ over the appropriate majorization polytope $\L(\vecx)$ or $\U(\vecx)$.

For non-uniformity,
\begin{equation}
\label{eq:prop1_nonuniformity}
\varepsilon^{\min}_{\L,d}
=
\inf_{\vecxp\in\L(\vecx)}
C(\vecxp,\vecy),
\qquad
\varepsilon^{\min}_{\U,d}
=
\inf_{\vecxp\in\U(\vecx)}
C(\vecxp,\vecy),
\end{equation}
where
\begin{equation}
\label{eq:prop1_objective_nonuniformity}
C(\vecx,\vecy)
=
\begin{cases}
g(\vecy-\vecx), & d=d_g,\\[1mm]
h\left(D_f(\vecx\|\vecy)\right), & d=S.
\end{cases}
\end{equation}

For pure-state entanglement and coherence,
\begin{equation}
\label{eq:prop1_entanglement_coherence}
\varepsilon^{\min}_{\L,d_g}
=
\inf_{\vecxp\in\U(\vecx)}
C(\vecxp,\vecy),
\qquad
\varepsilon^{\min}_{\U,d_g}
=
\inf_{\vecxp\in\L(\vecx)}
C(\vecxp,\vecy),
\end{equation}
where
\begin{equation}
\label{eq:prop1_objective_entanglement_coherence}
C(\vecx,\vecy)
=
g\left(
D_{\RI}(\vecx,\vecy),
-D_{\RI}(\vecx,\vecy),
\vec{0}_{D-2}
\right).
\end{equation}
\end{proposition}

\begin{proof}
We first establish a general reduction that depends only on the equivalence classes generated by reversible free transformations, and then identify the corresponding classical objective and majorization polytope for each resource theory.

Let $G$ be the unitary group implementing reversible free transformations associated with the corresponding MQRT. 
For each representative vector $\vecxp$ in the image of $\R$, choose an arbitrary state $\tau_{\vecxp}$ satisfying $\R(\tau_{\vecxp})=\vecxp$.
By construction of the representative map, it follows
\begin{equation}
\label{eq:representative_orbit}
\left\{
\tau\in\S:\R(\tau)=\vecxp
\right\}
=
\left\{
U\tau_{\vecxp}U^\dagger:U\in G
\right\}.
\end{equation}

Suppose that the distinguishability measure satisfies an orbit-reduction property of the form
\begin{equation}
\label{eq:reduction_measures_property}
\inf_{U\in G}
d\left(U\rho U^\dagger,\sigma\right)
=
C(\vecx,\vecy),
\end{equation}
where $\vecx=\R(\rho)$, $\vecy=\R(\sigma)$, and $C$ depends only on the representative probability vectors.

Let
\begin{equation}
X=\R\left(\L(\rho)\right),
\qquad
Y=\R\left(\U(\rho)\right).
\end{equation}
Using Eq.~\eqref{eq:representative_orbit}, the optimization over the lower conversion set decomposes as
\begin{align}
\varepsilon^{\min}_{\L,d}
&=
\inf_{\tau\in\L(\rho)}
d(\tau,\sigma)
\nonumber\\
&=
\inf_{\vecxp\in X}
\inf_{\R(\tau)=\vecxp}
d(\tau,\sigma)
\nonumber\\
&=
\inf_{\vecxp\in X}
\inf_{U\in G}
d\left(
U\tau_{\vecxp}U^\dagger,\sigma
\right).
\end{align}
Applying Eq.~\eqref{eq:reduction_measures_property}, we obtain
\begin{equation}
\label{eq:general_classical_reduction}
\varepsilon^{\min}_{\L,d}
=
\inf_{\vecxp\in X}
C(\vecxp,\vecy).
\end{equation}
Analogously,
\begin{equation}
\label{eq:general_classical_reduction_upper}
\varepsilon^{\min}_{\U,d}
=
\inf_{\vecxp\in Y}
C(\vecxp,\vecy).
\end{equation}
It remains to identify $X$, $Y$, and $C$ in each case.

\medskip

\noindent
\textbf{Non-uniformity.}

For non-uniformity, the representation map is order-preserving,
\begin{equation}
\rho\free\tau
\iff
\R(\rho)\succeq\R(\tau).
\end{equation}
Therefore,
\begin{equation}
\label{eq:nonuniformity_XY}
X=\L(\vecx),
\qquad
Y=\U(\vecx).
\end{equation}

For Schur-convex spectral distances $d_g$, Eq.\eqref{eq:quantum_Schur_convex_dist}, optimization over the unitary orbit gives \cite{Li2016}
\begin{equation}
\label{eq:spectral_orbit_nonuniformity}
\inf_{U\in U(D)}
d_g\left(U\rho U^\dagger,\sigma\right)
=
g(\vecy-\vecx).
\end{equation}
Hence,
\begin{equation}
C(\vecx,\vecy)=g(\vecy-\vecx).
\end{equation}

For the quantum divergences considered here, Eq.~\eqref{eq:quantum_divergence_reduction} directly gives
\begin{equation}
C(\vecx,\vecy)
=
h\left(D_f(\vecx\|\vecy)\right).
\end{equation}
Thus, the non-uniformity problems reduce to Eq.~\eqref{eq:prop1_nonuniformity} with the objectives in Eq.~\eqref{eq:prop1_objective_nonuniformity}.

\medskip

\noindent
\textbf{Pure-state entanglement and coherence.}

For pure-state entanglement and coherence, the representation map is order-reversing,
\begin{equation}
\rho\free\tau
\iff
\R(\tau)\succeq\R(\rho).
\end{equation}
Consequently,
\begin{equation}
\label{eq:entanglement_coherence_XY}
X=\U(\vecx),
\qquad
Y=\L(\vecx).
\end{equation}

Let
$\rho=\ket{\psi}\bra{\psi}$ and
$\sigma=\ket{\phi}\bra{\phi}$.
The difference of two pure states has at most two nonzero eigenvalues, given by
\begin{equation}
\pm\sqrt{1-|\braket{\psi|\phi}|^2}.
\end{equation}
Since Schur-convexity implies permutation invariance of $g$, we may therefore write
\begin{equation}
\label{eq:quantum_symmetric_pure}
d_g(\rho,\sigma) = g\left(r,-r,\vec{0}_{D-2}\right), \qquad r=\sqrt{1-|\braket{\psi|\phi}|^2}.
\end{equation}

Define
\begin{equation}
q_g(t) = g(t,-t,\vec{0}_{D-2}), \qquad 0\leq t\leq1.
\end{equation}
If $0\leq s\leq t$, then
\begin{equation}
(t,-t,\vec{0}_{D-2}) \succeq (s,-s,\vec{0}_{D-2}).
\end{equation}
Since $g$ is Schur-convex, it follows that
\begin{equation}
q_g(t)\geq q_g(s).
\end{equation}
Thus, $q_g$ is non-decreasing.

Now fix a representative vector $\vecxp$ and a pure state
$\tau_{\vecxp}=\ket{\psi_{\vecxp}}\bra{\psi_{\vecxp}}$
with $\R(\tau_{\vecxp})=\vecxp$.
Optimization over the reversible free orbit satisfies
\begin{equation}
\label{eq:fidelity_orbit_reduction}
\max_{U\in G}
\left|
\braket{\phi|U|\psi_{\vecxp}}
\right|
=
F(\vecxp,\vecy),
\end{equation}
for both bipartite pure-state entanglement and pure-state coherence.\footnote{The corresponding optimal-overlap reduction was established for bipartite pure-state entanglement in Ref.~\cite{Vidal2000} and for pure-state coherence in Ref.~\cite{Liu2022}.}
Equivalently,
\begin{equation}
\label{eq:RI_orbit_reduction}
\min_{U\in G}
\sqrt{
1-
\left|
\braket{\phi|U|\psi_{\vecxp}}
\right|^2
}
=
\sqrt{1-F^2(\vecxp,\vecy)}
=
D_{\RI}(\vecxp,\vecy).
\end{equation}

Using Eq.~\eqref{eq:quantum_symmetric_pure} and the fact that $q_g$ is non-decreasing, Eq.~\eqref{eq:RI_orbit_reduction} yields
\begin{align}
\inf_{U\in G}
d_g\left(U\tau_{\vecxp}U^\dagger,\sigma\right)
&=
q_g\left(D_{\RI}(\vecxp,\vecy)\right)
\nonumber\\
&=
g\left(D_{\RI}(\vecxp,\vecy),-D_{\RI}(\vecxp,\vecy),\vec{0}_{D-2}\right).
\end{align}
Hence the classical objective is Eq.\eqref{eq:prop1_objective_entanglement_coherence}, that is,
\begin{equation}
C(\vecx,\vecy)
=
g\left(D_{\RI}(\vecx,\vecy),-D_{\RI}(\vecx,\vecy),\vec{0}_{D-2}\right).
\end{equation}

Combining this orbit reduction with
Eqs.~\eqref{eq:general_classical_reduction},
\eqref{eq:general_classical_reduction_upper}, and
\eqref{eq:entanglement_coherence_XY}
gives Eq.~\eqref{eq:prop1_entanglement_coherence}, completing the proof.
\end{proof}

\section{Taut Strings: Definitions and Basic Properties}
\label{sec:taut_preliminaries}
We recall the main definitions and elementary properties of taut strings in a general discrete setting. Throughout this section, the admissible curves are continuous piecewise linear functions constrained to lie between two prescribed discrete boundaries.

\begin{definition}[Set of admissible strings]
Let
\begin{equation}
t_0<t_1<\cdots<t_n    
\end{equation}
be a set of nodes in $\mathbb R$, and let $\vec F,\vec G\in\mathbb R^{n+1}$ satisfy
\begin{equation}
F_k\le G_k,\qquad k=0,\ldots,n,
\end{equation}
with $F_0=G_0$ and $F_n=G_n$.

We define the set of admissible strings $\GammaFG$ as the set of continuous piecewise linear functions on $[t_0,t_n]$, with nodes at $t_0,\ldots,t_n$, and boundary vectors $\vec F$ (floor) and $\vec G$ (roof). That is,
\begin{equation}
\GammaFG
=
\bigl\{
\gamma\in C([t_0,t_n]) :
\gamma|_{[t_k,t_{k+1}]}\text{ is linear and }
F_k\le \gamma(t_k)\le G_k,\ k=0,\ldots,n
\bigr\}.
\end{equation}
\end{definition}

Equivalently, an admissible string is determined by its values at the nodes. It is often useful to describe such a string through its successive increments, which encode the vertical displacement of the curve between consecutive nodes.

\begin{definition}[Sets of admissible increments and slopes]
\label{def:Omega_FG}
Let $\GammaFG$ be the set of admissible strings associated with the nodes
$t_0<\cdots<t_n$ and the boundary vectors $\vec F$ and $\vec G$.
The corresponding set of admissible increments is defined by
\begin{equation}
\label{eq:Omega_FG}
\OmegaFG = \left\{\vecu \in \mathbb R^n \ \middle|\ u_k=\gamma(t_k)-\gamma(t_{k-1}), \quad k=1,\ldots,n, \text{ for some } \gamma\in\GammaFG \right\}.
\end{equation}
Notice that $s_n(\vecu)=F_n-F_0=G_n-G_0$ for all $\vecu \in \OmegaFG$.

The corresponding set of admissible slopes is defined by
\begin{equation}
\label{eq:S_FG}
 \Lambda_{\vecF,\vecG} =\left\{\vec v\in\mathbb R^n \ \middle|\ v_k=\frac{\gamma(t_k)-\gamma(t_{k-1})}  {t_k-t_{k-1}}, \quad k=1,\ldots,n, \text{ for some } \gamma\in\GammaFG \right\}.
\end{equation}

\end{definition}

The taut string is the shortest admissible curve inside the strip determined by the lower and upper boundaries. In the present discrete setting, its length is simply the polygonal length obtained by summing the Euclidean lengths of the linear pieces between consecutive nodes.

\begin{definition}[Taut string]
\label{def:taut_string}
Let $\GammaFG$ be the set of admissible strings associated with the nodes
$t_0<\cdots<t_n$, and boundary vectors $\vec F$ and $\vec G$.

For $\gamma\in \GammaFG$, define its length by
\begin{equation}
\label{eq:length_curve}
\mathcal L(\gamma)
=
\sum_{k=1}^{n}
\sqrt{
(t_k-t_{k-1})^2
+
\bigl(\gamma(t_k)-\gamma(t_{k-1})\bigr)^2
}.
\end{equation}

A function $\taut \in \GammaFG$ is called the \emph{taut string} if it has the minimal length among all functions in $\GammaFG$. That is,
\begin{equation}
\label{eq:taut_string}
\taut
\in
\operatorname*{arg\,min}_{\gamma\in \GammaFG}
\mathcal L(\gamma).
\end{equation}
\end{definition}

The preceding minimization problem is well posed. Indeed, the endpoint conditions ensure that all admissible strings share the same initial and final values, while the constraints at the nodes define a closed and bounded feasible set in finite dimension. Therefore, the existence of a minimizer is guaranteed. Moreover, the strict convexity of the length functional in the successive increments implies uniqueness, as stated in the following remark.

\begin{remark}[Existence and uniqueness of the taut string]
\label{remark:unicidad_taut_string}
Let $t_0<\cdots<t_n$ be a set of nodes, and let
$\vec F,\vec G\in\mathbb R^{n+1}$ satisfy
\begin{equation}
F_k\le G_k,\qquad k=0,\ldots,n,
\end{equation}
with $F_0=G_0$ and $F_n=G_n$. Let $\GammaFG$ be the
corresponding set of admissible strings. Then there exists a unique taut string
$\taut\in \GammaFG$.
\end{remark}

\begin{definition}[Least concave majorant and greatest convex minorant]
Let
\begin{equation}
    \mathcal{P}
    =
    \left\{(t_k,z_k)\right\}_{k=0}^{m},
    \qquad
    t_0<t_1<\cdots<t_m,
\end{equation}
be an ordered collection of points. The \emph{least concave majorant} of $\mathcal{P}$, denoted by
$\operatorname{LCM}(\mathcal{P})$, is the pointwise smallest concave
function on $[t_0,t_m]$ lying above all the points of $\mathcal{P}$:
\begin{equation}
    \operatorname{LCM}(\mathcal{P})(t)
    =
    \inf
    \left\{
        g(t):
        g:[t_0,t_m]\to\mathbb{R}
        \text{ is concave and }
        g(t_k)\geq z_k
        \text{ for all }k
    \right\}.
\end{equation}

Similarly, the \emph{greatest convex minorant} of $\mathcal{P}$, denoted by
$\operatorname{GCM}(\mathcal{P})$, is the pointwise largest convex
function on $[t_0,t_m]$ lying below all the points of $\mathcal{P}$:
\begin{equation}
    \operatorname{GCM}(\mathcal{P})(t)
    =
    \sup
    \left\{
        g(t):
        g:[t_0,t_m]\to\mathbb{R}
        \text{ is convex and }
        g(t_k)\leq z_k
        \text{ for all }k
    \right\}.
\end{equation}
\end{definition}

In some particular configurations, the taut string can be identified directly with a classical envelope of one of the two boundaries. 
The following two remarks make this observation precise: the first one treats the case in which the greatest convex minorant of the upper boundary is admissible, while the second one considers the dual case in which the least concave majorant of the lower boundary is admissible.

\begin{remark}[Taut string and the greatest convex minorant]
\label{remark:taut_string_gcm}
Let $t_0<\cdots<t_n$ be a set of nodes, and let
$\vec F,\vec G\in\mathbb R^{n+1}$ satisfy
\begin{equation}
F_k\le G_k,\qquad k=0,\ldots,n,
\end{equation}
with $F_0=G_0$ and $F_n=G_n$. Let $\GammaFG$ be the
corresponding set of admissible strings. 
If $\GCM\left(\{(t_k,G_k)\}_{k=0}^n\right) \in \GammaFG$, then the taut string $\taut$ of $\GammaFG$ coincides with $\GCM\left(\{(t_k,G_k)\}_{k=0}^n\right)$.
\end{remark}

\begin{remark}[Taut string and the least concave majorant]
\label{remark:taut_string_lcm}
Let $t_0<\cdots<t_n$ be a set of nodes, and let
$\vec F,\vec G\in\mathbb R^{n+1}$ satisfy
\begin{equation}
F_k\le G_k,\qquad k=0,\ldots,n,
\end{equation}
with $F_0=G_0$ and $F_n=G_n$. Let $\GammaFG$ be the
corresponding set of admissible strings. 
If $\LCM\left(\{(t_k,F_k)\}_{k=0}^n\right) \in \GammaFG$, then the taut string $\taut$ of $\GammaFG$ coincides with $\LCM\left(\{(t_k,F_k)\}_{k=0}^n\right)$.
\end{remark}

\section{Schur-Convex Objectives: Geometries T1 and T2}
\label{sec:schur_taut}

We now specialize the general taut-string framework of Sec.~\ref{sec:taut_preliminaries} to the uniformly spaced nodes $\{t_k=k/n\}^n_{k=0}$ and relate it to Schur-convex optimization over the lower and upper majorization polytopes. 
This yields the two elementary constructions T1 and T2, which solve the corresponding lower- and upper-set problems for arbitrary Schur-convex objectives.

The following lemma provides the key structural property of taut strings on uniformly spaced nodes: their increment vector is minimal with respect to majorization among all admissible increments, together with the corresponding boundary-contact conditions. 
These facts will be the basic ingredients for the T1 and T2 constructions.
\begin{lemma}\label{lemma:min_omega}
Let $\{ t_k = k/n \}_{k= 0}^n$ be a set of nodes, and let
$\vec F,\vec G\in\mathbb R^{n+1}$ satisfy
\begin{equation}
F_k\le G_k,\qquad k=0,\ldots,n,
\end{equation}
with $F_0=G_0$ and $F_n=G_n$. Let $\GammaFG$ be the
 set of admissible strings and $\OmegaFG$ the set of admissible increments.

Then,  the following properties hold:
\begin{enumerate}

    \item There exists a unique increment vector $\vectau \in \OmegaFG$ that is majorized by every vector in $\OmegaFG$; that is 
    $\vecu \succeq \vectau$ for all $\vecu \in \OmegaFG$. Moreover,  the vector $\vectau$ is the increment vector of the taut string $\taut$ of $\GammaFG$, that is  $\tau_k = \taut(t_k) - \taut(t_{k-1})$ for $k= 1,\ldots, 
    n$.\label{item:min_omeg}

    \item If $\taut$ is strictly locally convex at $k \in \{1,\ldots,n-1\}$ , i.e., $\tau_k < \tau_{k+1}$, then $\taut(t_k) = G_k$. \label{item:min_omeg_techo}

    \item If $\taut$ is strictly locally concave at $k \in \{1,\ldots,n-1\}$, i.e., $\tau_k > \tau_{k+1}$, then $\taut(t_k) = F_k$. \label{item:min_omeg_piso}

\end{enumerate}
\end{lemma}

\begin{proof}

~

\begin{enumerate}
\item 

From Theorem~5.1 of Ref. \cite{Kruglyak2016}, we have that there is a slope vector $\vecv^*\in\Lambda_{\vecF, \vecG}$ such that
$\vecv \succeq \vecv^*$ for every $\vecv \in\Lambda_{\vecF, \vecG}$.
Then, $\vecu^* = \vecv^*/n \in\OmegaFG$ satisfies $\vecu \succeq \vecu^*$ for every $\vecu\in\OmegaFG$.

Let $\vectau$ be the increment vector of the taut string $\taut$ of $\GammaFG$. We are going to show that $\vecu^* = \vectau$. 
Since $\vectau \in\OmegaFG$, we have $\vectau \succeq \vecu^*$. 
Let $\ell:\mathbb{R}^n\to\mathbb{R}$ be defined by 
\begin{equation}
\ell(\vec x)=\sum_{k=1}^{n}\sqrt{\frac{1}{n^2}+x_k^2}.     
\end{equation} 
Since $\ell$ is symmetric and separable, and the function $t\mapsto \sqrt{1/n^2+t^2}$ is convex, $\ell$ is Schur-convex. 
Therefore, $\ell(\vectau) \geq \ell(\vecu^*)$. 

Let $\gamma^* \in \GammaFG$ be the curve associated with the increment vector $\vecu^*$.
We observe that $\ell(\vectau) = \L (\taut)$ and $\ell(\vecu^*) = \L (\gamma^*)$, with $\L(\taut)$ and  $\L (\gamma^*)$ the curve lengths of $\taut$ and $\gamma^*$, respectively.
Therefore, $\L(\taut)\geq \L(\gamma^*)$.

On the other hand, since $\taut$ is the taut string, it minimizes the length
over $\GammaFG$, and therefore $\L(\taut)\leq \L(\gamma^*)$

Thus equality holds, so $\gamma^*$ is also a length minimizer in $\GammaFG$. By the
uniqueness of the taut string, $\gamma^*=\taut$, and consequently
$\vecu^*=\vectau$.

\item 
    This statement follows, for instance, from \cite[Theorem~3.1(iii)]{Grasmair2008}; see also \cite[p.~544]{Dantzig1971}.

    \item 
    This statement follows, for instance, from \cite[Theorem~3.1(iii)]{Grasmair2008}; see also \cite[p.~544]{Dantzig1971}.

\end{enumerate}
\end{proof}

We now record the taut-string properties associated with the lower and upper sets. The first lemma treats the lower-set construction, while the second one gives its upper-set counterpart. 
Both results will be used to establish the main result of this section and, ultimately, Proposition \ref{prop:reduction_taut_string}.

\begin{lemma}[Geometry T1: lower majorization set]
\label{lemma:y-x_Lower_set}
Let $\{ t_k = k/n \}_{k= 0}^n$ be a set of nodes. Let $\vecx,\vecy \in \pset$. Associated with $\vecy - \vecx$, we define the floor  boundary vector $\vec F \in \Real^{n+1}$ and the roof boundary vector $\vec G \in \Real^{n+1}$ as follows 
\begin{equation}
\label{eq:FG_definition_lower_T1}
F_k = s_k(\vec y-\vec x),
\qquad
G_k =
\begin{cases}
0 & k=0,n,\\
1 & k=1,\ldots,n-1.
\end{cases}
\end{equation}
For these boundary vectors we consider the set of admissible strings 
$\GammaFG$, its corresponding taut string $\taut$, 
the set of admissible increments $\OmegaFG$, and the associated increment vector $\vectau \in \OmegaFG$ given by
\begin{equation}\label{eq:vector_incrementos_lower}
\tau_k = \taut(t_k) - \taut(t_{k-1}) \quad \text{for } k=1,\dots,n. 
\end{equation}
Analogously, for $\vecx '\in\L(\vecx)$ and $\vecy\in\pset$, we consider $\vec y - \vec x'$, and we introduce the associated boundary vectors
$\vecF',\vecG'$ and the following objects:
$\Gamma_{\vecF',\vecG'}$,
$\tautp$,
$\Omega_{\vecF',\vecG'}$, and
$\vectau'$.

Then, the following properties hold:

\begin{enumerate}

    \item $\taut$ is the least concave majorant of $\{(t_k, F_k)\}^n_{k=0}$.
   \label{item:Lower_taut_uce}
       
   \item $\vecy - \vectau \in \L(\vecx)$. \label{item:Lower_in_pset}
    \item $\vecy - \vecx \succeq \vectau$. \label{item:Lower_y-x>tau} 
    \item If $\vecx ' \in \L(\vecx)$, then $\vectau' \succeq \vectau$. \label{item:Lower_preserving_1}

\end{enumerate}
\end{lemma}

\begin{proof}

~

\begin{enumerate}

 \item Let $C_{\vec F} = \LCM\left(\{(t_k, F_k )\}^n_{k=0}\right)$, that is, the least concave majorant of $\{(t_k, F_k )\}^n_{k=0}$. By definition, $C_{\vec F}$ is a concave continuous piecewise linear, (with nodes $\{ t_k = k/n \}_{k= 0}^n$) such that $C_{\vec F}(t_k) \geq F_k =s_k(\yx)$, for $k = 0,  \ldots, n$. 
 Moreover, the linear interpolation of $\{(t_k,G_k)\}^n_{k=0}$ is concave continuous piecewise linear, such that $G_k \geq  s_k(\yx)$, for $k = 0,  \ldots, n$. Then, the least concave majorant satisfies $C_{\vec F}(t_k) \leq  G_k$ for $k = 0,  \ldots, n$.
Summing up, $F_k \leq C_{\vec F}(t_k) \leq G_k$ for $k = 0,  \ldots, n$, which implies $C_{\vec F} \in \GammaFG$. 
Therefore, by Remark \ref{remark:taut_string_lcm}, we conclude $C_{\vec F} = \taut$.

\item 

Let $\vecz = \vecy - \vectau $. We observe that $s_k(\vectau) = \taut(t_k)$  for $k= 0, \ldots, n$, which follows from $\vectau$ being the increment vector of $\taut$ and $\taut(t_0) = 0$. 

First, we prove that $\vec z \in \pset$; that is, its entries sum to one, are
non-increasing, and are nonnegative. 

\begin{enumerate}[label=(\roman*)]
\item \textit{Normalization property}  
   \begin{equation}
     s_n(\vecz) = s_n(\vecy) - s_n(\vectau) = s_n(\vecy) - \taut(t_n)= 1,   
    \end{equation}
where we used the facts that $s_n(\vec y)=1$ and $\taut(t_n)=0$.

\item \textit{Non-increasing property}

We consider the difference between consecutive components of $\vecz$:
\begin{equation}\label{eq:diferencias_vecz}
            z_k - z_{k+1} = (y_k - \tau_{k} )- (y_{k+1} - \tau_{k+1}), \quad \text{for } k=1,\dots,n-1.
\end{equation}

By statement \ref{item:Lower_taut_uce} of this Lemma, the taut string $\taut$ is the least concave majorant of $\{(t_k, s_k(\vecy-\vecx))\}^n_{k=0}$. Therefore, the increment vector $\vectau$ of $\taut$ satisfies $\tau_{k} \geq  \tau_{k+1}$ for $k=1,\dots,n-1$. There are two cases.

\begin{itemize}
    \item Case $\tau_{k} = \tau_{k+1}$:
    \begin{equation}
    z_k - z_{k+1} = y_k - y_{k+1} \geq 0,
    \end{equation}
    where we used the fact that $\vecy$ is  a non-increasing probability vector.

    \item Case $\tau_{k} > \tau_{k+1}$: We consider the following identities
\begin{equation}
\begin{alignedat}{3}
y_k
&= s_{k}(\vecy)  -s_{k-1}(\vecy),
\qquad&
\tau_{k}
&= \taut(t_k)-\taut(t_{k-1}),\\
y_{k+1}
&= s_{k+1}(\vecy)  -s_{k}(\vecy),
\qquad&
\tau_{k+1}
&= \taut(t_{k+1})-\taut(t_k),
\end{alignedat}
\end{equation}
Replacing these identities into Eq. \eqref{eq:diferencias_vecz} we obtain
\begin{equation}\label{eq:nonincreasing_case2_diferencia}
            z_k - z_{k+1} =  2 (s_{k}(\vecy) - \taut(t_k) ) + \taut(t_{k-1}) -s_{k-1}(\vecy)  +\taut(t_{k+1}) -  s_{k+1}(\vecy).
\end{equation}

Since $\tau_{k} > \tau_{k+1}$, by statement \ref{item:min_omeg_piso} of Lemma \ref{lemma:min_omega}, we have $\taut(t_k) = F_k = s_k(\vecy-\vecx)$, which implies 
\begin{equation}\label{eq:nonincreasing_case2_k}
s_{k}(\vecy)- \taut(t_k)= s_{k}(\vecx).
\end{equation}
Moreover, since the taut string is greater or equal than the lower boundary, $F_k = s_k(\vecy-\vecx)$, we have 
\begin{equation}\label{eq:nonincreasing_case2_k+1-1}
\taut(t_{k-1})-s_{k-1}(\vec y)\geq -s_{k-1}(\vec x),
\qquad
\taut(t_{k+1})-s_{k+1}(\vec y)\geq -s_{k+1}(\vec x).
\end{equation}
Using Eqs.~\eqref{eq:nonincreasing_case2_k} and
\eqref{eq:nonincreasing_case2_k+1-1} in
Eq.~\eqref{eq:nonincreasing_case2_diferencia}, we obtain
\begin{equation}
            z_k - z_{k+1} \geq  2 s_{k}(\vecx) - s_{k-1}(\vecx)   -  s_{k+1}(\vecx).
\end{equation}
Since $\vec x\in\pset$, its entries are non-increasing, and therefore the partial
sums of $\vec x$ are concave, which implies
\begin{equation}
    2s_k(\vec x) -s_{k-1}(\vec x)- s_{k+1}(\vec x) \geq 0.
\end{equation}
Therefore, we conclude that $z_k\geq z_{k+1}$.

\end{itemize}

\item \textit{Non-negativity property}  

Since the entries of $\vec z$ are non-increasing, its smallest entry is $z_n$.
Therefore, it suffices to prove that $z_n\geq 0$. We observe that
\begin{equation}\label{eq:sum_uce_vs_x}
            s_k(\vecz) =
              s_k(\vecy) - s_k(\vectau) = s_k(\vecy) - \taut(t_k) \leq s_k(\vecx) \quad \text{for } k=0,\dots,n,  
        \end{equation}
where we have used again the fact that the taut string is greater or equal than the lower boundary, \ie $\taut(t_k) \geq  s_k(\vecy-\vecx)$.
Then, 
 \begin{equation}
            z_n  =  s_{n}(\vecz) - s_{n-1}(\vecz)=
            1 - s_{n-1}(\vecz) \geq 1 - s_{n-1}(\vecx) = x_n \geq 0,
\end{equation}
where we used that $s_n(\vec z)=s_n(\vec x)=1$ and that $\vec x$ is nonnegative.
    
\end{enumerate}

Having established that $\vecz \in \pset$, Ineqs. \eqref{eq:sum_uce_vs_x} are equivalent to $\vecx \succeq \vecz$, and it implies $\vecz = \vecy - \vectau  \in \L(\vecx)$.

\item 
We observe that $\vec y-\vec x \in \OmegaFG$. Indeed, $\vec y-\vec x$ is the increment vector of the linear interpolation of $s_k(\yx)=F_k$, which belongs to $\GammaFG$. Therefore, by statement \ref{item:min_omeg} of Lemma \ref{lemma:min_omega}, we obtain $\vec y-\vec x \succeq \vectau $

\item 
Assume $\vecx ' \in \L(\vecx)$. This implies $s_k(\vecx) \geq s_k(\vecx ')$ for $k = 0, \ldots, n$. Then, $s_k(\vecy - \vecx ') \geq s_k(\vecy - \vecx)$ for $k = 0, \ldots, n$. Thus, the lower boundaries satisfy  $F'_k \geq F_k$ for $k = 0, \ldots, n$, while the upper  boundaries remain equal,  $\vecG = \vec G'$.
    By definition $G'_k \geq \tautp(t_k) \geq  F'_k  $ for $k = 0, \ldots, n$. 
    Then, $G_k\geq \tautp (t_k) \geq  F_k$ for $k = 0, \ldots, n$.
    Hence, $\tautp \in \GammaFG$, and its increment vector $\vectau ' \in \OmegaFG$. Then, 
    by statement \ref{item:min_omeg} of Lemma \ref{lemma:min_omega}, we have $\vectau '  \succeq \vectau$.
     
    \end{enumerate}
\end{proof}

\begin{lemma}[Geometry T2: upper majorization set]
\label{lemma:y-x_Upper_set}

Let $\{ t_k = k/n \}_{k= 0}^n$ be a set of nodes. Let $\vecx,\vecy \in \pset$.
Associated with $\vecy - \vecx$, we define the floor boundary vector $\vecF \in \Real^{n+1}$
and the roof boundary vector $\vecG \in \Real^{n+1}$ as follows 
\begin{equation}
\label{eq:FG_definition_upper_T2}
F_k =
\begin{cases}
0 & k=0,\\
 s_k(\vec y)-1 & k=1,\ldots,n,
\end{cases} \qquad G_k = s_k(\yx).
\end{equation}
For these boundary vectors we consider the set of admissible strings $\GammaFG$, its corresponding taut string $\taut$, 
the set of admissible increments $\OmegaFG$, and the associated increment vector $\vectau\in \OmegaFG$ given by
\begin{equation}\label{eq:vector_incrementos_upper}
\tau_k = \taut(t_k) - \taut(t_{k-1}) \quad \text{for } k=1,\dots,n. 
\end{equation}

Analogously, for $\vecx '\in\U(\vecx)$ and $\vecy\in\pset$, we consider $\vec y - \vec x'$, and we introduce the associated boundary vectors
$\vecF',\vecG'$ and the following objects:
$\Gamma_{\vecF',\vecG'}$,
$\tautp$,
$\Omega_{\vecF',\vecG'}$, and
$\vectau'$.

Then, the following properties hold:

\begin{enumerate}
   
    \item $\vecy - \vectau \in \U(\vecx)$. \label{item:Upper_in_pset}
    \item $\vecy - \vecx \succeq \vectau$. \label{item:Upper_y-x>tau} 
    \item If $\vecx ' \in \U(\vecx)$, then $\vectau' \succeq \vectau$. \label{item:Upper_preserving_1}
\end{enumerate}
\end{lemma}

\begin{proof}

~

\begin{enumerate}

\item 

Let $\vecz = \vecy - \vectau$. Firstly, we are going to show that the entries of $\vecz$ are non-increasing.
The difference between consecutive components of $\vecz$ is given by
\begin{equation}\label{eq: upper_diferencias_vecz}
            z_k - z_{k+1} = y_k - y_{k+1} +  \tau_{k+1}- \tau_{k} , \quad \text{for } k=1,\dots,n-1.
\end{equation}
We split the proof into two cases, according to whether the taut string
$\taut$ is convex or not.

\begin{enumerate}
\item Convex case:

In this case, $\taut$ is convex, then, $\tau_k\leq \tau_{k+1}$ for $k= 1,\ldots, n-1$. Then, the non-increasing property follows from Eqs. \eqref{eq: upper_diferencias_vecz} and the fact that $\vecy$ is a non-increasing vector.

\item Non-convex case: 

In this case, the set $    \left\{k\in\{1,\ldots,n-1\}:\tau_k>\tau_{k+1}\right\}$ is nonempty. Then,
\begin{equation}
k^*=\min\left\{k\in\{1,\ldots,n-1\}:\tau_k>\tau_{k+1}\right\}.  
\end{equation}

By statement \ref{item:min_omeg_piso} of Lemma \ref{lemma:min_omega}, $\taut(t_{k^*}) = F_{k^*}  = s_{k^*}(\vecy) - 1 \leq 0 $.

For $k=1,\dots,k^*-1$, the increment vector $\tau$ satisfies  $\tau_{k}  \leq \tau_{k+1}$. Then, 
\begin{equation}\label{eq: upper_nonincreasing_nonconvex_primero}
        z_k - z_{k+1} = y_k - y_{k+1} +  \tau_{k+1}- \tau_{k} \geq 0 \quad \text{for }  1 \leq k \leq k^*-1,
\end{equation}
where we used the fact that $\vecy$ is a non-increasing vector.

For $k=k^*,\dots, n$, the taut string satisfies $\taut(t_k) \geq s_k(\vecy) - 1$, and its endpoints satisfy  $\taut(t_{k^*}) = s_{k^*}(\vecy) - 1 $ and $\taut(t_n) = s_{n}(\vecy) - 1$. 
Since $s_k(\vecy) - 1$ is concave, applying the Remark  \ref{remark:taut_string_lcm} to the restricted problem in the interval $[t_k^*, t_n]$, we have 
\begin{equation}\label{eq: upper_nonpositive_nonconvex_segundo}
\taut(t_k) = s_k(\vecy) - 1    \quad \text{for }  k^* \leq k \leq n.   
\end{equation}
This implies, for $k =k^*+1, \ldots, n$, that $\tau_k= y_k$.
Then,
\begin{equation}\label{eq: upper_nonincreasing_nonconvex_segundo}
        z_k - z_{k+1} = y_k - y_{k+1} +  \tau_{k+1}- \tau_{k} = 0 \quad \text{for }  k^* +1 \leq k \leq n-1.
\end{equation}

Finally, for $k = k^*$:
\begin{equation}\label{eq: upper_nonincreasing_nonconvex_tercero}
\begin{aligned}
z_{k^*}-z_{k^*+1}
&=
y_{k^*}-y_{k^*+1}
+\tau_{k^*+1}-\tau_{k^*}
=
y_{k^*}-\bigl(\taut(t_{k^*})-\taut(t_{k^*-1})\bigr)\\
&=
y_{k^*}-\bigl(s_{k^*}(\vec y)-1\bigr)+\taut(t_{k^*-1})
=
\taut(t_{k^*-1})- F_{k^*-1} \geq 0.
\end{aligned}
\end{equation}

Therefore, from Eqs. \eqref{eq: upper_nonincreasing_nonconvex_primero}, \eqref{eq: upper_nonincreasing_nonconvex_segundo}, and \eqref{eq: upper_nonincreasing_nonconvex_tercero}, we conclude that $z_{k}-z_{k+1} \geq 0$ for $k=1,\ldots,n-1 $.

\end{enumerate}

Secondly, we prove the normalization property:
\begin{equation}
     s_n(\vecz) = s_n(\vecy) - s_n(\vectau)= 1,   
\end{equation}
where we used the facts that 
$s_n(\vec y)=1$ and $s_n(\vectau) = 0$.

Finally, to prove nonnegative property, it suffices to prove that $z_n\geq 0$. We observe that
\begin{equation}
            z_n = y_n -   \tau_{n} =   y_n - (\taut(t_n)-\taut(t_{n-1}) )     \geq y_n + s_{n-1}(\vecy) - 1 = 0,
\end{equation}
where we used the fact that $\taut(t_n) = 0$ and $\taut(t_{n-1}) \geq F_{n-1} = s_{n-1}(\vecy) - 1$.

To verify $\vecz \in \U(\vecx)$, we use the upper boundary constraint, $\taut(t_k) \leq G_k = s_k(\vecy-\vecx)$. Equivalently, 
$s_k(\vecx) \leq s_k(\vecy) - \taut(t_k) = s_k(\vecz)$, which means $\vecz \succeq \vecx$. Therefore, $\vecz \in \U(\vecx)$.

\item 

 We observe that $\vec y-\vec x \in \OmegaFG$. Indeed, $\vec y-\vec x$ is the increment vector of the linear interpolation of $s_k(\yx)=G_k$, which belongs to $\GammaFG$. Therefore, by statement \ref{item:min_omeg} of Lemma \ref{lemma:min_omega}, we obtain $\vec y-\vec x \succeq \vectau$

     \item Assume $\vecx ' \in \U(\vecx)$. This implies $s_k(\vecx ') \geq s_k(\vecx)$ for $k = 0, \ldots, n$. Then, $s_k(\vecy - \vecx) \geq s_k(\vecy - \vecx ')$ for $k = 0, \ldots, n$. 
     Thus, the upper boundaries satisfy $G_k \geq G'_k$ for $k=0,\ldots,n$, while the lower boundaries remain equal, $\vecF = \vecF'$.
    By definition $G'_k\geq \tautp(t_k) \geq F'_k$ for $k = 0, \ldots, n$. Then, $G_k \geq \tautp(t_k) \geq  F_k$ for $k = 0, \ldots, n$. 
    Hence, $\tautp\in \GammaFG$, and its increment vector $\vectau' \in \OmegaFG$. 
    Then, by statement \ref{item:min_omeg} of Lemma \ref{lemma:min_omega}, we have $\vectau'  \succeq \vectau$.

\end{enumerate}
\end{proof}

We now state the main result of this section, which connects Schur-convex optimization over lower and upper sets with the taut-string constructions introduced above.

\begin{lemma}[Optimal Schur-convex function approximation via taut strings]
\label{lemma:lower_upper_inf}
Let $\vecx,\vecy\in\pset$, and let $\varphi:\Real^n\to\Real$ be Schur-convex.
Then the following statements hold.

\begin{enumerate}[label=\textup{(\roman*)}, leftmargin=0.8cm]

\item Lower set case (geometry T1):
\begin{equation}
    \inf_{\vecx'\in\L(\vecx)} \varphi(\vecy-\vecx') = \varphi(\vectau^{(1)}),
\end{equation}
and the infimum is attained at
\begin{equation}
\label{eq:argmin_lower}
    \vec y- \vectau^{(1)} \in \L(\vecx),
\end{equation}
where $\vectau^{(1)}$ denotes the increment vector of the taut string $\taut$ of the set $\GammaFG$, where the set of nodes is $\{ t_k = k/n \}_{k= 0}^n$, 
the floor $\vecF$ and the roof $\vecG$  are defined as
\begin{equation}
\label{eq:FG_definition_lower}
F_k =s_k(\vec y-\vec x),
\qquad
G_k=
\begin{cases}
0 & k=0,n,\\
1 & 1\le k\le n-1.
\end{cases}
\end{equation}

\item Upper set case (geometry T2):

\begin{equation}
    \inf_{\vec x ' \in\U(\vec x)} \varphi(\vec y-\vec x ')    =    \varphi(\vectau^{(2)}),
\end{equation}
and the infimum is attained at
\begin{equation}
    \vec y- \vectau^{(2)},
\end{equation}
where $\vectau^{(2)}$ denotes the increment vector of the taut string
$\taut$ of the set $\GammaFG$, where the set of nodes is $\{ t_k = k/n \}_{k= 0}^n$, the floor $\vecF$ and the roof $\vecG$  are defined as
\begin{equation}
\label{eq:FG_definition_upper}
F_k =
\begin{cases}
0 & k=0,\\
 s_k(\vec y)-1 & k=1,\ldots,n,
\end{cases}
\qquad
G_k = s_k(\vec y - \vec x).
\end{equation}

\end{enumerate}
\end{lemma}

\begin{proof}

~

\begin{enumerate}[label=\textup{(\roman*)}, leftmargin=0.8cm]

\item Lower set case:

Let $\vec x,\vec y \in \pset$. 
For any $\vec x ' \in \L(\vec x)$, consider the pair of vectors $\vec x', \vec y$, and let $\vectau'$ denote the increment vector of the taut string
$\tautp$ of $\Gamma_{\vec F', \vec G'}$, where the  boundary vectors $\vecF '$ and $\vecG '$ are defined as in Eq.\eqref{eq:FG_definition_lower} using $\vec x'$ and $\vecy$ instead of $\vec x$ and $\vec y$. 

Then, by statements
\ref{item:Lower_y-x>tau} and \ref{item:Lower_preserving_1} of Lemma~\ref{lemma:y-x_Lower_set}, we have that $\vec y-\vec x' \succeq  \vectau '    \succeq \vectau $.
Since $\varphi$ is Schur-convex, 
\begin{equation}
    \varphi(\vec y-\vec x ') \geq \varphi(\vectau),
    \qquad \forall\,\vec x ' \in \L(\vec x).
\end{equation}
On the other hand, by statement~\ref{item:Lower_in_pset} of Lemma~\ref{lemma:y-x_Lower_set},
$\vec x ^*=\vec y- \vectau$ belongs to $\L(\vec x)$.
At $\vec x ^*$ the lower bound is attained, since $\varphi\bigl(\vec y-(\vec y-\vectau)\bigr) =   \varphi(\vectau)$.

Consequently,
\begin{equation}
    \inf_{\vec x' \in \L(\vec x)} \varphi(\vec y-\vec x')  = \varphi(\vectau),
    \qquad
    \vec y-\vectau     \in \operatorname*{arg\,min}_{\vec x' \in \L(\vec x)} \varphi(\vec y-\vec x').
\end{equation}

\item Upper set case:

The proof follows mutatis mutandis by invoking Lemma~\ref{lemma:y-x_Upper_set} in place of Lemma~\ref{lemma:y-x_Lower_set}.

\end{enumerate}
\end{proof}

The following example illustrates the relevant case of $p$-norm objectives.
In particular, for the $1$-norm, the lower-set formula recovers the result originally obtained in \cite{Torgersen1989} (and later re-derived
in \cite{Horodecki2018}), whereas the corresponding upper-set formula recovers the result obtained in \cite{Zanoni2024}.

\begin{example}[$p$-norm objective functions]
\label{ex:p_norms}
For $p\in[1,\infty]$, the function
\begin{equation}
\varphi(\vec z)=\|\vec z\|_p
\end{equation}
is Schur-convex. Hence, applying
Lemma~\ref{lemma:lower_upper_inf} to geometry T1 gives
\begin{equation}
\label{eq:p_norm_lower}
\min_{\vecxp\in\L(\vecx)}
\|\vecy-\vecxp\|_p
=
\|\vectau^{(1)}\|_p,
\end{equation}
where $\vectau^{(1)}$ is the increment vector of the T1 taut string.

For $p=1$, Eq.~\eqref{eq:p_norm_lower} reduces to the closed-form expression
\begin{equation}
\label{eq:inf_norm_1_lower}
\min_{\vecxp\in\L(\vecx)}
\|\vecy-\vecxp\|_1
=
2\max_{1\leq k\leq n-1}
\left\{
0,s_k(\vecy-\vecx)
\right\},
\end{equation}
which recovers the result originally derived in \cite{Torgersen1989} (and later rederived in ~\cite{Horodecki2018}).

Similarly, geometry T2 yields
\begin{equation}
\label{eq:p_norm_upper}
\min_{\vecxp\in\U(\vecx)}
\|\vecy-\vecxp\|_p
=
\|\vectau^{(2)}\|_p,
\end{equation}
where $\vectau^{(2)}$ is the increment vector of the T2 taut string.

For $p=1$, Eq.~\eqref{eq:p_norm_upper} becomes
\begin{equation}
\label{eq:inf_norm_1_upper}
\min_{\vecxp\in\U(\vecx)}
\|\vecy-\vecxp\|_1
=
2\max_{1\leq k\leq n-1}
\left\{
0,-s_k(\vecy-\vecx)
\right\},
\end{equation}
recovering the upper-set result of \cite{Zanoni2024}.
\end{example}

\section{$\vecd$-Majorization and $f$-Divergences: Definitions and Basic Properties}
\label{sec:d-maj}

To address the $f$-divergence objectives associated with geometries T3 and T4, we use the relative version of majorization known as $\vecd$-majorization. In our application, the reference vector will be the target distribution, $\vecd=\vecy$. We follow Refs.~\cite{marshall2011inequalities,Renes2016}.

\begin{definition}[$\vecd$-majorization]
Let $\vecx,\vecy,\vecd\in\pset$. We say that $\vecx$ $\vecd$-majorizes $\vecy$, and write
$\vecx\succeq_{\vecd}\vecy$, if there exists a column-stochastic matrix $D$ such that
\begin{equation}
D\vecd=\vecd,
\qquad
D\vecx=\vecy.
\end{equation}
\end{definition}

When $\vecd$ is the uniform distribution, $\vecd=(1/n,\ldots,1/n)$, the condition $D\vecd=\vecd$ implies that $D$ is doubly stochastic. Hence, $\vecd$-majorization reduces to ordinary majorization.

Following Ref.~\cite{Renes2016}, we introduce the Lorenz-diagram representation of $\vecd$-majorization.

\begin{definition}[$\vecd$-Lorenz diagram]
Given $\vecx,\vecd\in\pset$, the $\vecd$-Lorenz diagram of $\vecx$ is
\begin{equation}
\mathfrak T(\vecx\mid\vecd)
=
\left\{
\left(
\sum_{i=1}^{n}a_i d_i,
\sum_{i=1}^{n}a_i x_i
\right)
:
0\leq a_i\leq1
\right\}.
\end{equation}
\end{definition}

Although $\vec x$ and $\vec d$ are written in non-increasing order, the boundaries of $\mathfrak{T}(\vec x\mid\vec d)$ are determined by the ordering of the ratios $x_i/d_i$.
These boundaries admit a simple polygonal representation in terms of the ratios $x_i/d_i$ \cite{Renes2016}.

\begin{lemma}[Ordered-ratio representation of the $\vecd$-Lorenz boundaries]
\label{lemma:d_lorenz_ordered_ratios}
Let $\sigma$ be a permutation that orders the ratios $x_i/d_i$ in non-increasing order. Then the upper boundary of
$\mathfrak T(\vecx\mid\vecd)$ is the polygonal line joining the accumulated points
\begin{equation}
\left(
\sum_{i=1}^{k}d_{\sigma(i)},
\sum_{i=1}^{k}x_{\sigma(i)}
\right),
\qquad
k=0,\ldots,n.
\end{equation}
Similarly, if $\pi$ orders the ratios $x_i/d_i$ in non-decreasing order, the lower boundary is the polygonal line joining
\begin{equation}
\left(
\sum_{i=1}^{k}d_{\pi(i)},
\sum_{i=1}^{k}x_{\pi(i)}
\right),
\qquad
k=0,\ldots,n.
\end{equation}
Whenever $d_i=0<x_i$, we set $x_i/d_i=+\infty$. Indices for which $x_i=d_i=0$ may be placed arbitrarily, since they do not contribute to the accumulated points.
\end{lemma}

\begin{definition}[$\vecd$-Lorenz curves]
The upper and lower $\vecd$-Lorenz curves, denoted by
$\mathfrak U_{\vecx\mid\vecd}$ and
$\mathfrak L_{\vecx\mid\vecd}$, respectively, are the functions whose graphs are the upper and lower boundaries of
$\mathfrak T(\vecx\mid\vecd)$. For $c\in[0,1]$,
\begin{equation}
\mathfrak U_{\vecx\mid\vecd}(c)
=
\max
\left\{
t:(c,t)\in\mathfrak T(\vecx\mid\vecd)
\right\},
\end{equation}
and
\begin{equation}
\mathfrak L_{\vecx\mid\vecd}(c)
=
\min
\left\{
t:(c,t)\in\mathfrak T(\vecx\mid\vecd)
\right\}.
\end{equation}
\end{definition}

\begin{remark}
The upper $\vecd$-Lorenz curve $\mathfrak U_{\vecx\mid\vecd}$ is concave, whereas the lower curve
$\mathfrak L_{\vecx\mid\vecd}$ is convex. Moreover,
$\mathfrak T(\vecx\mid\vecd)$ is centrally symmetric with respect to $(1/2,1/2)$. Indeed, replacing $a_i$ by $1-a_i$ maps any point $(c,t)$ of the diagram to $(1-c,1-t)$. Consequently,
\begin{equation}
\mathfrak L_{\vecx\mid\vecd}(c)
=
1-
\mathfrak U_{\vecx\mid\vecd}(1-c),
\qquad
c\in[0,1].
\end{equation}
\end{remark}

We now recall the equivalent Lorenz-diagram characterizations of $\vecd$-majorization \cite{Renes2016,Blackwell1953}.

\begin{theorem}[Characterizations of $\vecd$-majorization]
\label{thm:d_majorization_characterization}
Let $\vecx,\vecy,\vecd\in\pset$. The following statements are equivalent:
\begin{enumerate}[label=\textup{(\roman*)}, leftmargin=0.8cm]
\item
$\vecx\succeq_{\vecd}\vecy$;

\item
$\mathfrak T(\vecy\mid\vecd)
\subseteq
\mathfrak T(\vecx\mid\vecd);
$

\item
$\mathfrak U_{\vecy\mid\vecd}(c) \leq \mathfrak U_{\vecx\mid\vecd}(c),$ for all $c\in[0,1]$;

\item
$\mathfrak L_{\vecy\mid\vecd}(c) \geq \mathfrak L_{\vecx\mid\vecd}(c)$, for all  $c\in[0,1]$.

\end{enumerate}
\end{theorem}

A natural family of monotones for $\vecd$-majorization is provided by Csisz\'ar $f$-divergences \cite{Csiszar1967}.

\begin{definition}[Csisz\'ar $f$-divergence]
Let $f:\mathbb R_{+}\to\Real$ be convex. The $f$-divergence of $\vecx$ relative to $\vecd$ is
\begin{equation}
D_f(\vecx\Vert\vecd)
=
\sum_{i=1}^{n}
d_i
f\left(\frac{x_i}{d_i}\right),
\end{equation}
where terms with $d_i=0$ are understood in the extended sense
\begin{equation}
0\,f\left(\frac{x_i}{0}\right)
=
\begin{cases}
0, & x_i=0,\\[1mm]
x_i f_{\infty}, & x_i>0,
\end{cases}
\qquad \text{with} \ 
f_{\infty}
=
\lim_{u\to\infty}\frac{f(u)}{u}.
\end{equation}

\end{definition}

Csisz\'ar $f$-divergences satisfy the data-processing inequality and are therefore monotone under $\vecd$-majorization \cite{Ruch1978}. In particular,
\begin{equation}
\vecx\succeq_{\vecd}\vecy
\quad\Longrightarrow\quad
D_f(\vecy\Vert\vecd)
\leq
D_f(\vecx\Vert\vecd).
\label{eq:f_divergence_d_majorization_monotonicity}
\end{equation}

\section{$f-$divergence Objectives: Geometries T3 and T4}
\label{sec:fdiv_taut}

Throughout this section, $\vecy$ is fixed and its dependence is suppressed from the notation. 
We set $m=|\supp(\vecy)|\leq n$.

For geometries T3 and T4, the set of nodes is $\{t_k = s_k(\vecy)\}^m_{k=0}$, and the relevant order is $\vecy$-majorization. 
We first construct suitable lower and upper $\vecy$-Lorenz curves from greatest convex minorants and least concave majorants, respectively, and then use the monotonicity
of $f$-divergences under $\vecy$-majorization.

\begin{definition}[Greatest convex-minorant family]
Let $\vecx,\vecy\in\pset$ and $m=|\supp(\vecy)|\leq n$. 
For each $\vecz\in\L(\vecx)$ satisfying $s_m(\vecz)=s_m(\vecx)$, 
we denote by $\eta_{\vecz}$ the greatest convex minorant of the
polygonal curve determined by the points $\left\{
\bigl(s_k(\vecy),s_k(\vecz)\bigr)
\right\}_{k=0}^{m}$.
Equivalently,
\begin{equation}
\eta_{\vecz}
=
\operatorname{GCM}
\left(
\left\{
\bigl(s_k(\vecy),s_k(\vecz)\bigr)
\right\}_{k=0}^{m}
\right).
\end{equation}

We then define the set of all such greatest convex minorants by
\begin{equation}
\Theta(\vecx,m)
=
\left\{
\eta_{\vecz}
:
\vecz\in\L(\vecx)
\text{ and }
s_m(\vecz)=s_m(\vecx)
\right\},
\end{equation}
and the corresponding set of increment vectors by $\Omega_{\Theta}(\vecx,m)$.
\end{definition}

\begin{lemma}[Geometry T3: lower majorization set]
\label{lemma: m lower set f divergence}
Let $\vecx,\vecy\in\Delta_n^\downarrow$, with $|\supp(\vecy)|=m\leq n$.  
For $k=0,\ldots,m$, let $
    t_k=s_k(\vec y)$ be a set of nodes.
Associated with $\vecx$ and $\vecy$, 
let us consider the sets $\Theta(\vecx,m)$ and $\Omega_{\Theta}(\vecx,m)$, and the floor and roof boundary
vectors $\vec F,\vec G\in\mathbb R^{m+1}$ defined as follows
\begin{equation}
    F_k=\begin{cases}
        0&\text{if    } k=0,\ldots,m-1\\
        s_m(\vecx)&\text{if } k=m
    \end{cases}
    \qquad
    G_k=s_k(\vecx),\qquad k=0,\ldots,m.
\end{equation}
For these boundary vectors we consider the set of admissible strings $\Gamma_{\vec F,\vec G}$, its corresponding taut string $\taut$, the set
of admissible increments $\OmegaFG$ , and the associated increment vector $\vec{\tau}\in \OmegaFG$ given by
\begin{equation}
    \tau_k
    =
    \taut(s_k(\vecy))-\taut(s_{k-1}(\vecy)),
    \qquad k=1,\ldots,m.
\end{equation}

Then the following properties hold:
\begin{enumerate}
    \item \label{item:T3_item1} The set of extended vectors $\widehat{\Omega}_{\Theta}(\vecx,m) =\Big\{\vec{w}\oplus(x_{m+1},\ldots,x_n)\,:\, \vec{w}\in\Omega_{\Theta}(\vecx,m)\big\}$ is included in $\L(\vec x)$. 
    
    \item \label{item:T3_item2}
     The taut string $\taut$ is the greatest convex minorant of the points $\left\{\bigl(s_k(\vecy),s_k(\vecx)\bigr)\right\}_{k=0}^{m}$, \ie
     \begin{equation}
    \taut =\eta_{\vecx} ~~~with ~~~\eta_{\vecx}  = \GCM\left(\left\{\bigl(s_k(\vecy),s_k(\vecx)\bigr)\right\}_{k=0}^{m}\right).       
     \end{equation}

 \item \label{item:T3_item3} For every $\vecx' \in\L(\vecx)$, the increment vector $\vec{w} \in \Omega_{\Theta}(\vecx',m)$ of $\eta_{\vecx'}  = \GCM\left(
\left\{\bigl(s_k(\vecy),s_k(\vecx')\bigr)\right\}_{k=0}^{m}\right)$  satisfies
    \begin{equation}
    \vecx'\succeq_{\vecy}   \vec{w}\oplus (x'_{m+1},\ldots, x'_n).
    \end{equation}

\end{enumerate}
\end{lemma}

\begin{proof}

~

\begin{enumerate}
        \item Let $\widehat{\vec{w}} = \vec{w}\oplus(x_{m+1},\ldots,x_n) \in \widehat{\Omega}_{\Theta}(\vecx,m)$. First we show that  $\widehat{\vec{w}}\in \pset$.
          Since $\vec{w} \in \Omega_{\Theta}(\vecx,m)$, there is a $\vec z\in\L(\vec x)$ with $s_m(\vecz)=s_m(\vecx)$ such that $\eta_{\vecz}$ has $\vec{w}$ as an increment vector. 

\textbf{Normalization property:} if we compute the sum of the entries, we have 
\begin{align}
\sum_{k=1}^n \widehat{w}_k=\sum_{k=1}^m w_k + \sum_{k=m+1}^n x_k.
\end{align}
      
Since $w_k=\eta_{\vec z}(s_k(\vecy))-\eta_{\vec z}(s_{k-1}(\vecy))$, and $\eta_{\vec z}(s_0(\vecy))=s_0(\vecz)=0$ and $\eta_{\vec z}(s_m(\vecy))=s_m(\vecx)$, we have
\begin{equation}
\sum_{k=1}^n \widehat{w}_k =\eta_{\vec z}(s_m(\vecy))- \eta_{\vec z}(s_0(\vecy)) + \sum_{k=m+1}^n x_k =s_m(\vecx) + \sum_{k=m+1}^n x_k = s_n(\vecx)= 1.
\end{equation}

\textbf{Non-increasing property:} 

\textit{Case $k > m$:}
\begin{equation}
    \widehat{w}_{k} -  \widehat{w}_{k+1} = x_{k} -  x_{k+1} \ge 0, 
\end{equation}
where we have used that $\vecx \in \pset $.

\textit{Case $k = m$:} we have $\widehat{w}_{m} -  \widehat{w}_{m+1} = w_{m} -  x_{m+1}$.

We observe that $\eta_{\vec{z}}(s_m(\vecy))=s_m(\vecx)$ and
\begin{equation}
\eta_{\vec{z}}(s_{m-1}(\vecy)) \leq s_{m-1}(\vecz) \leq s_{m-1}(\vecx). 
\end{equation}
The first inequality is due to the definition of $\eta_{\vec{z}}$, the second inequality follows from $\vecz\in\L(\vecx)$. 
Therefore, 
\begin{align}
    w_m&=\eta(s_m(\vecy))-\eta(s_{m-1}(\vecy))\\
    &\ge s_m(\vecx)-s_{m-1}(\vecx)=x_m\\
    &\ge x_{m+1}.
\end{align}
Therefore, $\widehat{w}_{m} -  \widehat{w}_{m+1} = w_{m} -  x_{m+1} \geq 0$.

\textit{Case $k < m$:} we have $\widehat{w}_{k} -  \widehat{w}_{k+1} = w_{k} -  w_{k+1}$.

There are two sub-cases: $\eta_{\vec z}(s_k(\vecy))=s_k(\vecz)$ or $\eta_{\vec z}(s_k(\vecy))<s_k(\vecz),$

Sub-case $\eta_{\vec z}(s_k(\vecy))=s_k(\vecz)$:

Since $\eta_{\vec z}(s_{k-1}(\vecy))\leq s_{k-1}(\vecz)$ and
$\eta_{\vec z}(s_{k+1}(\vecy))\leq s_{k+1}(\vecz)$, we get
\begin{equation}
    w_k
    =
    \eta_{\vec z}(s_k(\vecy))-\eta_{\vec z}(s_{k-1}(\vecy))
    \geq
    s_k(\vecz)-s_{k-1}(\vecz)
    =
    z_k,
\end{equation}
and
\begin{equation}
    w_{k+1}
    =
    \eta_{\vec z}(s_{k+1}(\vecy))-\eta_{\vec z}(s_k(\vecy))
    \leq
    s_{k+1}(\vecz)-s_k(\vecz)
    =
    z_{k+1}.
\end{equation}
Since $\vec z\in\Delta_n^\downarrow$, then $ z_k
    \geq
    z_{k+1}$. This implies
\begin{equation}
    w_k
    \geq
    z_k
    \geq
    z_{k+1}
    \geq
    w_{k+1}.
\end{equation}

Sub-case $\eta_{\vec z}(s_k(\vecy)) < s_k(\vecz)$:

Let us consider the adjacent slopes of the intervals $[k-1,k]$ and $[k,k+1]$:
\begin{equation}
    \frac{\eta_{\vec z}(s_k(\vecy)) -\eta_{\vec z}(s_{k-1}(\vecy)) }{s_k(\vecy)- s_{k-1}(\vecy)}
    =\frac{w_k}{y_k}, ~~~ ~~~\frac{\eta_{\vec z}(s_{k+1}(\vecy)) -\eta_{\vec z}(s_{k}(\vecy)) }{s_{k+1}(\vecy)- s_{k}(\vecy)}
    =\frac{w_{k+1}}{y_{k+1}}.
\end{equation}
Notice that the slopes are non-negative.

As a consequence of $\eta_{\vec z}$ being the greatest convex minorant and the fact that $\eta_{\vec z}(s_k(\vecy)) < s_k(\vecz)$, the two adjacent slopes
coincide (see Lemma 1 of \cite{Francu2017}):
\begin{equation}
\frac{w_k}{y_k}
    =
    \frac{w_{k+1}}{y_{k+1}}.
\end{equation}
Since $y_k\geq y_{k+1} > 0$ and $w_{k+1}/y_{k+1} \geq 0$, we obtain
\begin{equation}
    w_k
    =
    \frac{y_k}{y_{k+1}}w_{k+1}
    \geq
    w_{k+1}.
\end{equation}

Therefore, $\widehat{w}_{k} -  \widehat{w}_{k+1} = w_{k} -  w_{k+1} \geq 0$.

\textbf{Non-negativity property:} if $m<n$, it follows from non-increasing property and $x_n \ge 0$. 
If $m=n$, it follows from non-increasing property and $w_m \geq 0$.

Therefore,  $\widehat{\vec{w}} \in  \pset$.

To prove  $\widehat{\vec{w}} \in \L(\vecx)$, we observe that 
\begin{equation}
    s_k(\widehat{\vec{w}})
    =
   \begin{cases}
        \eta_{\vec z}(s_k(\vec y)),
        & k=1,\ldots,m,\\
        s_k(\vec x),
        & k=m,\ldots,n.
    \end{cases}
\end{equation}

Since, for $k = 1, \ldots, m$, $\eta_{\vec z}(s_k(\vec y)) \leq s_k(\vec z)\leq s_k(\vec x)$, we have that $s_k(\widehat{\vec{w}}) \leq s_k(\vecx)$. 
Therefore, $\widehat{\vec{w}} \in \L(\vecx)$.

     \item Let $\eta_{\vecx}\in\Theta(\vecx,m)$ be the greatest convex minorant of the points $\left\{\bigl(s_k(\vecy),s_k(\vecx)\bigr)\right\}_{k=0}^{m}$. Since $F_k \leq \eta_{\vecx}(s_k(\vecy)) \leq G_k$, for $k= 0, \ldots, m$, we have $ \eta_{\vecx}\in\GammaFG$. Then, by Remark \ref{remark:taut_string_gcm}, we have that $\taut=\eta_{\vecx}$.

\item Let $\vecx' \in\L(\vec x)$ and define
\begin{equation}
\eta_{\vecx' }
=
\operatorname{GCM}\left(
\left\{
\bigl(s_k(\vec y),s_k(\vecx')\bigr)
\right\}_{k=0}^{m}\right)\in\Theta(\vecx',m).
\end{equation}
Let $\vec w\in\Omega_{\Theta}(\vecx' ,m)$ be its corresponding increment vector, and set
\begin{equation}
\widehat{\vecw}
=
\vec w
\oplus
(x'_{m+1},\ldots,x'_n).
\end{equation}
By item \ref{item:T3_item1} of this lemma, $\widehat{\vecw} \in\L(\vecx' )$. 

Moreover, since $\mathfrak L_{\vecx'\mid\vec y}$ is convex and 
\begin{equation}
\mathfrak L_{\vecx' \mid\vec y}
(s_k(\vec y)) \le s_k(\vecx' ), \qquad k=0,\ldots,m, 
\end{equation} 
it is a convex minorant of the points $\left\{\bigl(s_k(\vec y),s_k(\vecx')\bigr)\right\}_{k=0}^{m}$. 
Hence, by maximality of the greatest convex minorant, 
\begin{equation}
\mathfrak L_{\vecx' \mid\vec y}(c) \le
\eta_{\vecx'}(c), \qquad c\in[0,1]. 
\end{equation}
By construction of lower $\vecy$-Lorenz curve, $\mathfrak L_{\widehat{\vecw}\mid\vec y} = \eta_{\vecx' }$. 
Consequently,
\begin{equation}
\mathfrak L_{\vecx' \mid\vec y}(c)
\le
\mathfrak L_{\widehat{\vecw}\mid\vec y}(c),
\qquad c\in[0,1].
\end{equation}
By Theorem \ref{thm:d_majorization_characterization}, we have
\begin{equation}
\vecx'\succeq_{\vecy} \widehat{\vecw} = \vec w \oplus (x'_{m+1},\ldots,x'_n).
\end{equation}

\end{enumerate}
\end{proof}

\begin{lemma}[Geometry T4: upper majorization set]
\label{lemma: uper set f divergence}
Let $\vec x,\vec y\in\Delta_n^\downarrow$, with $|\mathrm{supp}(\vecy)|=m\le n$. For $k=0,\ldots,m$, let $t_k=s_k(\vec y)$ be a set of nodes. Associated with $\vec x$ and $\vec y$, define the floor and roof boundary
vectors $\vec F,\vec G\in\mathbb R^{m+1}$ by
\begin{equation}
    F_k=\begin{cases}
    s_k(\vec x),&k=0,\ldots,m-1,\\
    1&k=m, 
    \end{cases}
    \qquad
    G_k=
    \begin{cases}
        0, & k=0,\\
        1, & k=1,\ldots,m.
    \end{cases}
\end{equation}
For these boundary vectors we consider the set of admissible strings $\Gamma_{\vec F,\vec G}$, its corresponding taut string $\taut$, the set
of admissible increments $\OmegaFG$ , and the associated increment vector $\vec{\tau}\in \OmegaFG$ given by
\begin{equation}
    \tau_k
    =
    \taut(s_k(\vecy))-\taut(s_{k-1}(\vecy)),
    \qquad k=1,\ldots,m.
\end{equation} 

Also, we define
\begin{equation}
    \Delta_{\vec F,\vec G}=
    \{\gamma\in\Gamma_{\vec F,\vec G}:\gamma\text{ is concave}\},
\end{equation}
and its corresponding set of
increment vectors $\Omega_{\Delta_{\vec F,\vec G}}$.

Then the following properties hold:
\begin{enumerate}
    \item \label{item:T4_item1} The set of extended vectors 
    $\widehat{\Omega}_{\Delta_{\vecF,\vecG}}=\Big\{\vec{v}\oplus\vec{0}_{n-m}\,:\, \vec{v}\in \Omega_{\Delta_{\vecF,\vecG}} \big\}$ is included in $\U(\vecx)$.

    \item \label{item:T4_item2}
    The taut string $\taut$ is the least concave majorant of the points $ \left\{ \bigl(s_k(\vec y),s_k(\vecx )\bigr) \right\}_{k=0}^{m-1} \cup {(1,1)}$, \ie
\begin{equation} 
  \taut = \gamma_{\vecx} ~~~with ~~~\gamma_{\vecx} =
\operatorname{LCM} \left( \left\{ \bigl(s_k(\vec y),s_k(\vecx )\bigr) \right\}_{k=0}^{m-1} \cup {(1,1)} \right).
\end{equation}

\item \label{item:T4_item3} For every $\vecx'\in\U(\vecx)$, the increment vector $\vec{v}$ of $\gamma_{\vecx'}= 
\operatorname{LCM} \left( \left\{ \bigl(s_k(\vec y),s_k(\vecx'  )\bigr) \right\}_{k=0}^{m-1} \cup {(1,1)} \right)$ satisfies
    \begin{equation}
\vecx'\succeq_{\vecy}        \vec v\oplus\vec{0}_{n-m}.
    \end{equation}
    
  \end{enumerate}
\end{lemma}

\begin{proof}

~

\begin{enumerate}
\item Let $\widehat{\vec{v}}=\vec{v}\oplus \vec{0}_{n-m}\in \widehat{\Omega}_{\Delta_{\vec F,\vec G}}$. 
First we show that $\widehat{\vec{v}}\in\pset$. Since $\vec v\in \Omega_{\Delta_{\vecF,\vecG}}$, there exists $\gamma\in \Delta_{\vecF,\vecG}$ whose increment vector is $\vec v$. 

\textbf{Normalization property:}  if we compute the sum of the entries, we have
\begin{equation} 
\sum_{k=1}^{n}\widehat{v}_k= \sum_{k=1}^m v_k.
\end{equation}
Since $v_k=\gamma(s_k(\vecy))-\gamma(s_{k-1}(\vecy))$, with $\gamma(s_0(\vecy))=0$ and $\gamma(s_m(\vecy))=1$, we have
\begin{equation}  
\sum_{k=1}^m v_k=\sum_{k=1}^{m}\gamma(s_k(\vecy))-\gamma(s_{k-1}(\vecy))= \gamma(s_m(\vecy))- \gamma(s_0(\vecy))= 1.
\end{equation}

\textbf{Non increasing property:} 

\textit{Case $k > m$:} Since $\widehat{v}_{k} = \widehat{v}_{k+1} = 0$, the inequality $\widehat{v}_{k} \geq \widehat{v}_{k+1}$ holds trivially.

\textit{Case $k = m$:} We have
\begin{equation}
\widehat{v}_{m} = v_{m} = \gamma(s_m(\vec y)) - \gamma(s_{m-1}(\vec y)).    
\end{equation}

Since $\gamma(s_m(\vec y)) = 1$ and $\gamma(s_{m-1}(\vec y)) \leq 1$, it follows that $\widehat{v}_{m} \geq 0$.
Therefore, since $\widehat{v}_{m+1} = 0$, we have that $\widehat{v}_{m} \geq \widehat{v}_{m+1}$.

\textit{Case $k < m$:} Let us consider the adjacent slopes of the intervals $[k-1,k]$ and $[k,k+1]$:
\begin{equation}
    \frac{\gamma(s_k(\vecy)) -\gamma(s_{k-1}(\vecy)) }{s_k(\vecy)- s_{k-1}(\vecy)}
    =\frac{v_k}{y_k}, ~~~ ~~~\frac{\gamma(s_{k+1}(\vecy)) -\gamma(s_{k}(\vecy)) }{s_{k+1}(\vecy)- s_{k}(\vecy)}
    =\frac{v_{k+1}}{y_{k+1}}.
\end{equation}
Notice that the slopes are non-negative. 
In addition, since $\gamma$ is concave, its successive slopes are non-increasing, \ie $\frac{v_{k}}{y_{k}} \geq \frac{v_{k+1}}{y_{k+1}}$. 
Hence, since $\vec y\in\Delta_n^\downarrow$, we have $v_{k}\geq {v_{k+1}}  \frac{y_{k}}{y_{k+1}} \ge v_{k+1}$.

\textbf{Non-negativity property:} if $m<n$, it follows from non-increasing property  and $\widehat{v_n} = 0$. 
If $n=m$, it follows from non-increasing property  and $v_m \geq 0$.

Therefore, we conclude that $\widehat{\vecv}\in\pset$.

To prove $\widehat{\vecv}\in\U(\vecx)$, we observe that

\begin{equation} s_k(\widehat{\vecv}) = 
\begin{cases}
    \gamma(s_k(\vecy)),& k=1,\ldots,m-1\\
    1&k=m,\ldots,n
\end{cases}\end{equation} 
Since, for $k=1,\ldots,m-1$, $\gamma(s_k(\vecy))\ge F_k= s_k(\vecx)$, we have that $s_k(\widehat{\vecv})\geq s_k(\vecx)$, for $k=1,\ldots,n$. 
Then $\widehat{\vecv} \in \U(\vecx)$. 
This implies $\widehat{\Omega}_{\Delta_{\vecF,\vecG}} \subseteq \U(\vecx)$.

\item Let $\gamma_{\vec x}$ be the least concave majorant of the points  $ \left\{ \bigl(s_k(\vec y),s_k(\vecx )\bigr) \right\}_{k=0}^{m-1} \cup {(1,1)}$. Since $F_k \leq \gamma_{\vecx}(s_k(\vecy)) \leq G_k$, for $k= 0, \ldots, m$, we have $ \gamma_{\vecx}\in\GammaFG$. Then, by Remark \ref{remark:taut_string_lcm}, we have that $\taut=\gamma_{\vecx}$.

\item Let $\vecx' \in\U(\vec x)$ and 
\begin{equation} 
\gamma_{\vecx' } =
\operatorname{LCM} \left( \left\{ \bigl(s_k(\vec y),s_k(\vecx' )\bigr) \right\}_{k=0}^{m-1} \cup {(1,1)} \right).
\end{equation}

Since $\gamma_{\vecx' } \in \GammaFG$ and is concave, we have $\gamma_{\vecx'} \in \Delta_{\vec F,\vec G}$. 
Then, its increment vector $\vecv$ belongs to $\Omega_{\Delta_{\vecF,\vecG}}$ and $\widehat{\vecv} =\vecv \oplus \vec{0}_{n-m} \in \widehat{\Omega}_{\Delta_{\vecF,\vecG}}$.
By item \ref{item:T4_item1} of this lemma, $\widehat{\vecv} \in \U(\vecx)$. 

Moreover, since $\mathfrak U_{\vecx' \mid\vec y}$ is concave and 
\begin{equation}
s_k(\vecx')\le \mathfrak U_{\vecx' \mid\vec y}
(s_k(\vec y)) , \qquad k=0,\ldots,m, 
\end{equation} 
it is a concave majorant of the points $\left\{
\bigl(s_k(\vec y),s_k(\vecx' )\bigr)
\right\}_{k=0}^{m}$
Hence, by
minimality of the least concave majorant,
\begin{equation}
\gamma_{\vecx' }(c)\le \mathfrak U_{\vecx' \mid\vec y}(c), \qquad c\in[0,1]. \end{equation}
By construction of upper $\vecy$-Lorenz curve, $\mathfrak U_{\widehat{\vecv}\mid\vec y}= \gamma_{\vecx' }$. 
Consequently,
\begin{equation}
\mathfrak U_{\widehat{\vecv}\mid\vec y}(c)\leq \mathfrak U_{\vecx' \mid\vec y}(c),
\qquad c\in[0,1].
\end{equation}
By Theorem \ref{thm:d_majorization_characterization}, 
\begin{equation}
\vecx'\succeq_{\vecy}  \widehat{\vecv} = \vecv \oplus \vec 0_{n-m}
\end{equation}

\end{enumerate}
\end{proof}

\begin{lemma}[Optimal $f$-divergence approximation via taut strings]
\label{lemma:lower_upper_inf_diverg}
Let $\vecx,\vecy\in\pset$, with $|\mathrm{supp}(\vecy)|= m \leq n$, and $D_f$ be a Csiszár $f$-divergence.
Then the following statements hold.

\begin{enumerate}[label=\textup{(\roman*)}, leftmargin=0.8cm]

\item Lower set case (geometry T3):

\begin{equation}
\label{eq:inf_lower_divergence}
     \inf_{\vec x ' \in\L(\vec x)} D_f(\vecx'||\vecy) =D_f(\widevectau^{(3)}||\vecy),
\end{equation}
and the infimum is attained at
\begin{equation}
    \widevectau^{(3)}= \vectau^{(3)} \oplus (x_{m+1}, \ldots, x_n) \in \L(\vec x),
\end{equation}
where $\vectau^{(3)}$ denotes the increment vector of the taut string
$\taut$ of the set $\GammaFG$, where the set of nodes is given by $\{t_k=s_k(\vecy)\}^m_{k=0}$, and the floor $\vecF \in \mathbb R^{m+1}$ and the roof $\vecG  \in \mathbb R^{m+1}$ are defined as
\begin{equation}
    F_k=\begin{cases}
        0&\text{if    } k=0,\ldots,m-1\\
        s_m(\vecx)&\text{if } k=m
    \end{cases}
    \qquad
    G_k=s_k(\vecx),\qquad k=0,\ldots,m.
\end{equation}
\item Upper set case (geometry T4):
\begin{equation}
\label{eq:inf_upper_divergence}
    \inf_{\vec x ' \in\U(\vec x)} D_f(\vecx'||\vecy)   =    D_f(\widevectau^{(4)}||\vecy)
\end{equation}
and the infimum is attained at
\begin{equation}
\widevectau^{(4)}= \vectau^{(4)} \oplus\vec{0}_{n-m} \in \U(\vec x),    
\end{equation}
where $\vectau^{(4)}$ denotes the increment vector of the taut string $\taut$ of the set $\GammaFG$, where the set of nodes is given by $\{t_k=s_k(\vecy)\}^m_{k=0}$, 
and the floor $\vecF$ and the roof $\vecG$ are defined as
\begin{equation}
\label{eq:FG_definition_upper_diverg}
    F_k=\begin{cases}
    s_k(\vec x),&k=0,\ldots,m-1,\\
    1&k=m, 
    \end{cases}
    \qquad
    G_k=
    \begin{cases}
        0, & k=0,\\
        1, & k=1,\ldots,m.
    \end{cases}
\end{equation}

\end{enumerate}
\end{lemma}

\begin{proof}

~

\begin{enumerate}[label=\textup{(\roman*)}, leftmargin=0.8cm]

\item Lower set case:

Let $\widevectau = \vec\tau\oplus(x_{m+1},\ldots,x_n)$ and let $\vecx' \in\L(\vec x)$.
In order to prove $D_f(\widevectau ||\vecy) \leq  D_f(\vecx'||\vecy)$, we are going to show that $\vecx'\succeq_{\vecy} \widevectau$.

By item \ref{item:T3_item3} of Lemma \ref{lemma: m lower set f divergence}, the increment vector $\vec w\in\Omega_{\Theta}(\vecx',m)$ of $\eta_{\vecx'} = \GCM\left( \left\{ \bigl(s_k(\vec y),s_k(\vecx')\bigr) \right\}
_{k=0}^{m}\right)$ satisfies
\begin{equation} 
\vecx'\succeq_{\vecy}  \widehat{\vecw} = \vec w \oplus (x'_{m+1},\ldots,x'_n). 
\end{equation}
Also, by construction of lower $\vecy$-Lorenz curve, we have 
\begin{equation} 
 \eta_{\vecx'}(c) = \mathfrak L_{\widehat{\vecw}\mid\vec y}(c),  \qquad c\in[0,1].
\end{equation}

By item \ref{item:T3_item2} of Lemma \ref{lemma: m lower set f divergence}, 
\begin{equation} 
\taut=  \eta_{\vecx}= \GCM\left( \left\{ \bigl(s_k(\vec y),s_k(\vec x)\bigr) \right\}_{k=0}
^{m}\right),
\end{equation} 
with $\vec{\tau}$ the associated increment vector, and by construction we have 
\begin{equation} 
\eta_{\vec x}(c) = \mathfrak L_{\widevectau \mid \vecy}(c),  \qquad c\in[0,1].
\end{equation}

Since  $\eta_{\vecx'} = \GCM\left( \left\{ \bigl(s_k(\vec y),s_k(\vecx')\bigr) \right\}_{k=0}^{m}\right)$, $\eta_{\vecx'}(s_k(\vecy)) \leq s_k(\vecx')$ for $k=0,\ldots,m$.
Moreover, the condition $\vecx' \in\L(\vec x)$ implies $s_k(\vecx') \leq s_k(\vecx)$ for $k= 0,\ldots, m$. 
Then, the curve $\eta_{\vecx'}$ is a convex minorant of the points $\left\{ \bigl(s_k(\vec y),s_k(\vecx )\bigr) \right\}_{k=0}^{m}$. 
Therefore, $\eta_{\vecx'} \leq \eta_{\vecx}$, which implies
\begin{equation}
\mathfrak L_{\widehat{\vecw} \mid\vec y}(c) \leq \mathfrak L_{\widevectau \mid\vec y}(c), \qquad c\in[0,1]. 
\end{equation}
By Theorem \ref{thm:d_majorization_characterization}, we have
\begin{equation}
\widehat{\vecw} \succeq_{\vecy} \widevectau. 
\end{equation}
Together with $\vecx'\succeq_{\vecy} \widehat{\vecw}$, transitivity gives 
\begin{equation} 
\vecx'\succeq_{\vecy} \widevectau, \qquad \forall\,\vecx' \in\L(\vec x).  
\end{equation}
By monotonicity of $f$-divergences under $\vecy$-majorization,
\begin{equation}
D_f(\widevectau \mid\mid \vecy) \leq D_f(\vecx' \mid\mid\vec y), \qquad \forall\,\vecx' \in\L(\vec x). 
\end{equation}
Therefore, 
\begin{equation} 
D_f(\widevectau \mid\mid\vec y) \leq \inf_{\vec x'\in\L(\vec x)} D_f(\vecx'\mid\mid\vec y). 
\end{equation}

On the other hand, by items \ref{item:T3_item1} and \ref{item:T3_item2} of Lemma \ref{lemma: m lower set f divergence}, $\widevectau\in\L(\vec x)$.  
Hence, 
\begin{equation}
\inf_{\vec x'\in\L(\vec x)} D_f(\vec x'\mid\mid\vecy) \le D_f(\widevectau\mid\mid\vec y). 
\end{equation}
Combining both inequalities, 
\begin{equation}
\inf_{\vec x'\in\L(\vec x)} D_f(\vec x'\mid\mid\vec y) = D_f\left( \vec\tau\oplus(x_{m+1},\ldots,x_n) \mid\mid \vec y \right),
\end{equation} 
and the infimum is attained at $\vec\tau\oplus(x_{m+1},\ldots,x_n)$.

\item Upper set case:

Let $\widevectau=\vec{\tau}\oplus \vec{0}_{n-m}$ and let $\vecx' \in\U(\vec x)$.
In order to prove $D_f(\widevectau ||\vecy) \leq  D_f(\vecx'||\vecy)$, we are going to show that $\vecx'\succeq_{\vecy} \widevectau$.

By item \ref{item:T4_item3} of Lemma \ref{lemma: uper set f divergence}, the increment vector $\vec v$ of $\gamma_{\vecx'} = \LCM\left( \left\{ \bigl(s_k(\vec y),s_k(\vecx' )\bigr) \right\}_{k=0}^{m-1} \cup {(1,1)} \right)$ satisfies
\begin{equation} 
\vecx'\succeq_{\vecy}  \widehat{\vecv} = \vecv \oplus \vec{0}_{n-m}. 
\end{equation}
Also, by construction of upper $\vecy$-Lorenz curve, we have 
\begin{equation} 
 \gamma_{\vecx'}(c) = \mathfrak U_{\widehat{\vecv} \mid\vec y}(c),  \qquad c\in[0,1].
\end{equation}

By item \ref{item:T4_item2} of Lemma \ref{lemma: uper set f divergence}, 
\begin{equation} 
\taut=  \gamma_{\vecx}= \operatorname{LCM} \left( \left\{ \bigl(s_k(\vec y),s_k(\vecx )\bigr) \right\}_{k=0}^{m-1} \cup {(1,1)} \right),
\end{equation} 
with $\vec{\tau}$ the associated increment vector, and by construction we have 
\begin{equation} 
\gamma_{\vec x}(c) = \mathfrak U_{\widevectau \mid\vec y}(c),  \qquad c\in[0,1].
\end{equation}

By definition of   $\gamma_{\vecx'}$, we have  $\gamma_{\vecx'}(s_k(\vecy)) \geq s_k(\vecx')$ for $k=0,\ldots,m$.
Moreover, the condition $\vecx' \in\U(\vec x)$ implies $s_k(\vecx') \geq s_k(\vecx)$ for $k= 0,\ldots, m$. 
Then, the curve $\gamma_{\vecx'}$ is a concave majorant of the points 
\begin{equation}
\left\{ \bigl(s_k(\vec y),s_k(\vecx )\bigr) \right\}_{k=0}^{m-1} \cup{(1,1)}.    
\end{equation}
 Therefore, $\gamma_{\vecx'} \geq \gamma_{\vecx}$, which implies
\begin{equation}
\mathfrak U_{\widehat{\vecv} \mid\vec y}(c) \geq \mathfrak U_{\widevectau\mid\vec y}(c), \qquad c\in[0,1]. 
\end{equation}

By Theorem \ref{thm:d_majorization_characterization}, we have
\begin{equation}
\widehat{\vecv}\succeq_{\vecy} \widevectau. 
\end{equation}
Together with $\vecx'\succeq_{\vecy} \widehat{\vecv}$, transitivity gives 
\begin{equation} 
\vecx'\succeq_{\vecy} \widevectau, \qquad \forall\,\vecx' \in\U(\vec x).  
\end{equation}

By monotonicity of $f$-divergences under $\vec y$-majorization,
\begin{equation}
D_f(\widevectau \mid\mid \vecy) \leq D_f(\vecx' \mid\mid\vec y), \qquad \forall\,\vecx' \in\U(\vecx). 
\end{equation}
Therefore, 
\begin{equation} 
D_f(\widevectau\mid\mid\vec y) \leq \inf_{\vec x'\in\U(\vecx)} D_f(\vecx'\mid\mid\vec y). 
\end{equation}

On the other hand, by items \ref{item:T4_item1} and \ref{item:T4_item2} of Lemma \ref{lemma: uper set f divergence}, $\widevectau \in\U(\vec x)$.  
Hence,
\begin{equation}
\inf_{\vec x'\in\U(\vec x)} D_f(\vec x'\mid\mid\vec y) \leq D_f
\left( \vec\tau\oplus\vec{0}_{n-m} \mid\mid \vec y \right).
\end{equation}

Combining both inequalities, 
\begin{equation}
\inf_{\vec x'\in\U(\vec x)} D_f(\vec x'\mid\mid\vec y) = D_f\left( \vec\tau\oplus\vec{0}_{n-m} \mid\mid \vec y \right).
\end{equation} 
and the infimum is attained at $\vec\tau\oplus\vec{0}_{n-m}$.

\end{enumerate}

\end{proof}

\section{Proof of Proposition 2: Reduction to Taut-String Geometries}
\label{sec:prop2}

We now combine the results of the previous sections to prove
Proposition~2 of the Letter. The four elementary constructions
T1--T4 are those listed in Table~I of the Letter.

For geometries T3 and T4, let
\begin{equation}
m=|\supp(\vecy)|.
\end{equation}
Since their taut strings are defined on the nodes
$t_k=s_k(\vecy)$, $k=0,\ldots,m$, their increment vectors have
dimension $m$. We therefore introduce the corresponding
$n$-dimensional extensions
\begin{equation}
\widevectau^{(3)}
=
\vectau^{(3)}
\oplus
(x_{m+1},\ldots,x_n),
\qquad
\widevectau^{(4)}
=
\vectau^{(4)}
\oplus
\vec{0}_{n-m}.
\label{eq:tau_extensions_prop2}
\end{equation}
When $m=n$, these extensions are trivial.

\begin{proposition}[Reduction to taut-string geometries]
\label{prop:reduction_taut_string}
Let $\vecx,\vecy\in\pset$ with  $m=|\supp(\vecy)| \leq n$.

\begin{enumerate}[label=\textup{(\roman*)}, leftmargin=*]

\item \label{item:prop2_schur} Schur-convex objectives. 

For every Schur-convex function $\varphi:\Real^n\rightarrow\Real$,
\begin{equation}
    \min_{\vecx'\in \L(\vecx)}   \varphi(\vecy-\vecx')    =    \varphi\!\left(\vec{\tau}^{(1)}\right),
\end{equation}
where $\vec{\tau}^{(1)}$ is the increment vector associated with geometry T1, and 
\begin{equation}
    \vecy-\vec{\tau}^{(1)}\in \L(\vecx)
\end{equation}
is a minimizer.

Similarly,
\begin{equation}
    \min_{\vecx'\in \U(\vecx)}    \varphi(\vecy-\vecx')     =    \varphi\!\left(\vec{\tau}^{(2)}\right),
\end{equation}
where $\vec{\tau}^{(2)}$ is the increment vector associated with geometry T2, and
\begin{equation}
    \vecy-\vec{\tau}^{(2)}\in \U(\vecx)
\end{equation}
is a minimizer.

\item \label{item:prop2_fdiv} $f$-divergence objectives.

Let $f:\mathbb{R}_{+}\rightarrow\mathbb{R}$ be convex, and let $D_f$ be the corresponding Csisz\'ar $f$-divergence.
Then
\begin{equation}
    \min_{\vecx'\in \L(\vecx)}
    D_f(\vecx'\Vert\vecy)    =    D_f\!\left(\widevectau^{(3)}  \Vert \vecy  \right),
\end{equation}
where $\vec{\tau}^{(3)}$ is the increment vector associated with geometry T3, and
\begin{equation}
    \widevectau^{(3)}    =   \vec{\tau}^{(3)} \oplus  (x_{m+1},\ldots,x_n) \in \L(\vecx)
\end{equation}
is a minimizer.

Similarly,
\begin{equation}
    \min_{\vecx'\in \U(\vecx)}    D_f(\vecx'\Vert\vecy) =    D_f\!\left(\widehat{\vec{\tau}}^{(4)}  \Vert\vecy   \right),
\end{equation}
where $\vec{\tau}^{(4)}$ is the increment vector associated with geometry T4, and
\begin{equation}
    \widevectau^{(4)}    =   \vec{\tau}^{(4)} \oplus \vec{0}_{n-m}  \in \U(\vecx)
\end{equation}
is a minimizer.
When $m=n$, the extensions are trivial.

\end{enumerate}

\end{proposition}

\begin{proof}
Statements~\ref{item:prop2_schur} and \ref{item:prop2_fdiv} follow directly from Lemmas~\ref{lemma:lower_upper_inf} and \ref{lemma:lower_upper_inf_diverg}, respectively.
In particular, the latter lemma yields the extended minimizers $\widevectau^{(3)}$ and $\widevectau^{(4)}$ when $m<n$.
\end{proof}


\section{Numerical Example}
\label{sec:example}

We illustrate the four taut-string constructions with a concrete
numerical example. Let $n=12$ and consider
\begin{align}
\vecx
&=
\frac{1}{5000}
(1364,\,1340,\,292,\,292,\,292,\,292,\,
282,\,282,\,200,\,180,\,100,\,84),
\\
\vecy
&=
\frac{1}{5000}
(951,\,951,\,762,\,752,\,752,\,240,\,
148,\,148,\,148,\,148,\,0,\,0).
\end{align}
Both vectors belong to $\pset$, with
\begin{equation}
|\supp(\vecx)|=12=n,
\qquad
|\supp(\vecy)|=m=10<n.
\end{equation}
Moreover, neither $\vecx\succeq\vecy$ nor
$\vecy\succeq\vecx$ holds, so both majorization directions lead to
nontrivial approximation problems. Since $\vecy$ does not have full
support, this example also illustrates explicitly the $n$-dimensional
extensions required for geometries T3 and T4.

The same pair $(\vecx,\vecy)$ can be used as the pair of representative
vectors for each of the three majorization-based quantum resource
theories considered in the Letter. Namely, we take initial and target
states $\rho$ and $\sigma$ satisfying
\begin{equation}
\R(\rho)=\vecx,
\qquad
\R(\sigma)=\vecy.
\end{equation}
For non-uniformity, $\vecx$ and $\vecy$ specify the spectra of the
states; for bipartite pure-state entanglement, their squared Schmidt
coefficients; and for pure-state coherence, their populations in the
fixed incoherent basis. Thus, the construction depends only on the
representative vectors and does not require the states to share an
eigenbasis or Schmidt basis.

\medskip
\noindent
\textit{Quantum and classical conversion sets.---}
For non-uniformity, the representation map is order preserving, so the
quantum and classical conversion directions coincide:
\begin{equation}
\varepsilon_{\L,d}^{\min}
=
\inf_{\vecxp\in\L(\vecx)}
C(\vecxp,\vecy),
\end{equation}
and analogously for $\U$.

For pure-state entanglement and coherence, instead, the representation
map is order reversing. Hence Proposition~\ref{prop:prop1} gives
\begin{equation}
\varepsilon_{\L,d_g}^{\min}
=
\inf_{\vecxp\in\U(\vecx)}
C(\vecxp,\vecy),
\qquad
\varepsilon_{\U,d_g}^{\min}
=
\inf_{\vecxp\in\L(\vecx)}
C(\vecxp,\vecy).
\end{equation}
Consequently, for entanglement and coherence, geometry T4 solves the
quantum optimization over $\L(\rho)$, whereas geometry T3 solves that
over $\U(\rho)$.

\medskip
\noindent
\textit{The four taut-string solutions.---}
For the fixed pair $(\vecx,\vecy)$, the four taut strings depend only on
the corresponding majorization constraints and not on the particular
distinguishability measure within each class covered by the main
theorem.

\begin{figure}[!htbp]
\centering
\includegraphics[height=0.55\linewidth]{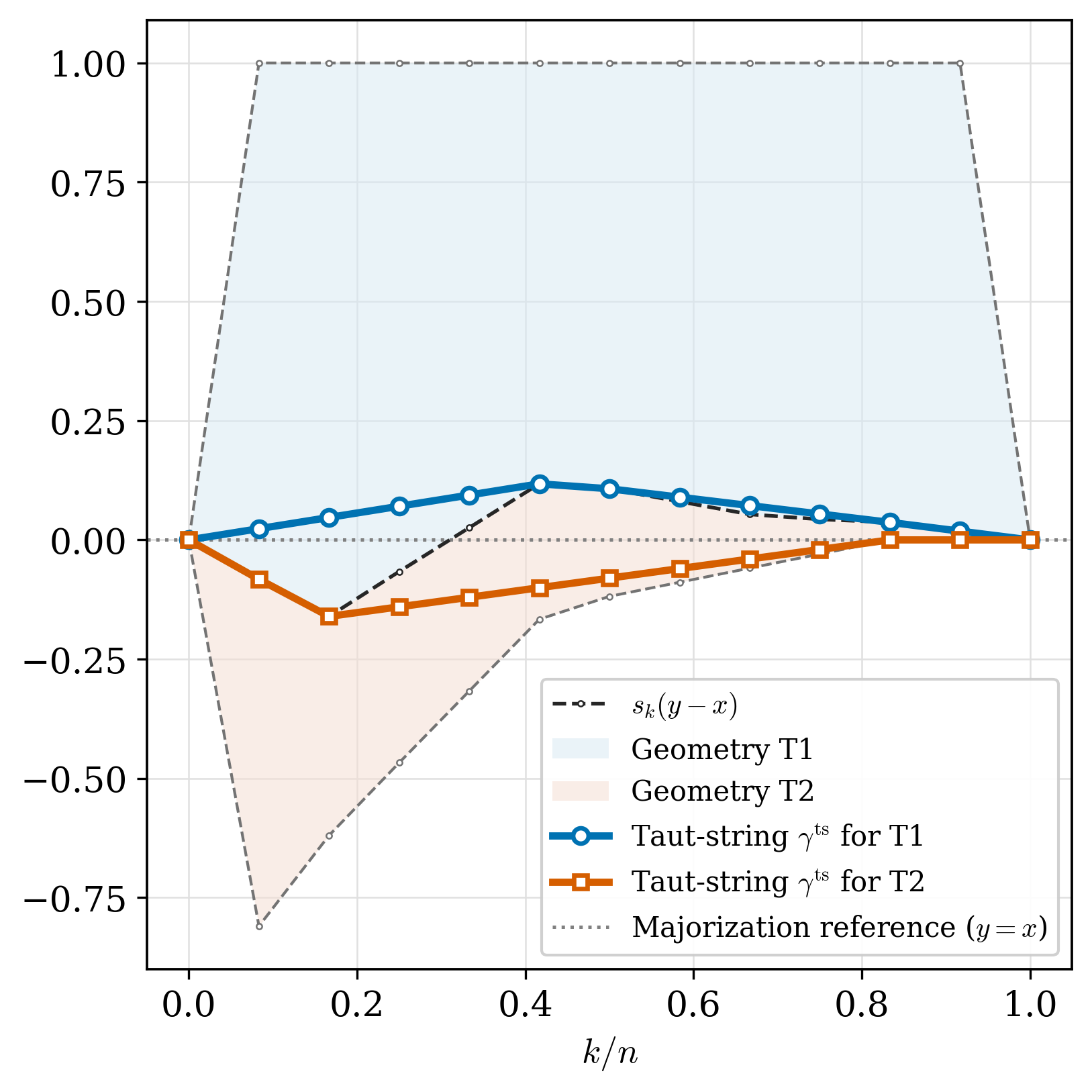}
\caption{
Taut strings T1 and T2 for the pair $(\vecx,\vecy)$ of
Sec.~\ref{sec:example}. They solve the lower- and
upper-majorization-polytope problems, respectively, for the
Schur-convex objectives $g(\vecy-\vecxp)$. Dashed lines denote the
floor and roof, with the boundary shared by T1 and T2 shown in black;
solid lines denote the corresponding taut strings.
}
\label{fig:t1t2}
\end{figure}

Geometries T1 and T2 solve the lower- and upper-polytope problems,
respectively, for objectives of the form
$g(\vecy-\vecxp)$ with $g$ Schur-convex. In the quantum setting, these
are precisely the classical objectives arising from spectral distances
for non-uniformity. Both constructions use the uniform grid
$t_k=k/n$. Figure~\ref{fig:t1t2} shows their floors, roofs, and taut
strings.

Geometries T3 and T4 solve the lower- and upper-polytope $f$-divergence problems, respectively, and use the generally non-uniform grid
\begin{equation}
t_k=s_k(\vecy), \qquad k=0,\ldots,m.
\end{equation}
As shown in the proof of the main theorem in the Letter, through the fidelity reduction the same two geometries also solve the pure-state entanglement and
coherence objectives generated by spectral distances.

\begin{figure}[!htbp]
\centering
\includegraphics[height=0.55\linewidth]{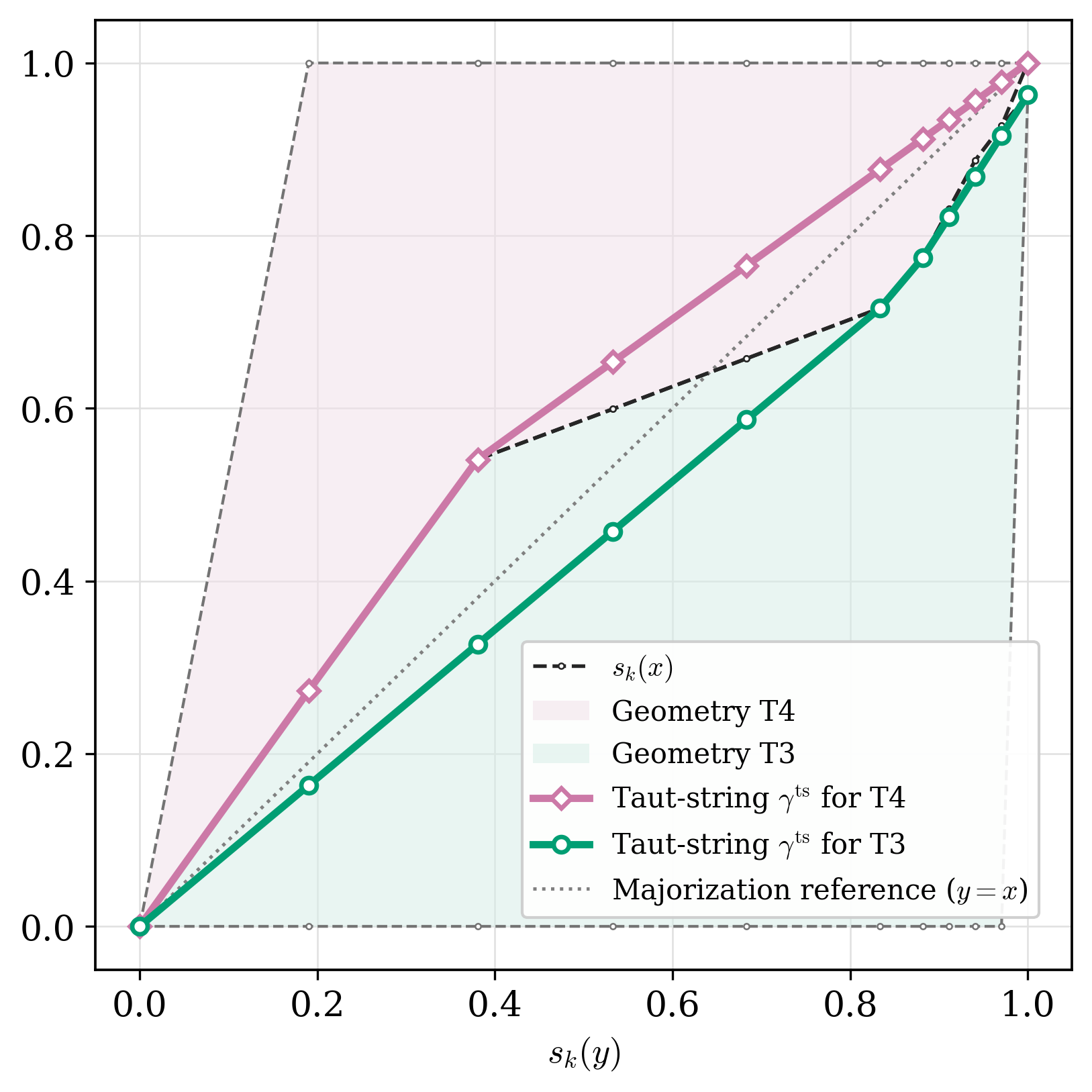}
\caption{
Taut strings T3 and T4 for the same pair $(\vecx,\vecy)$. They solve
the lower- and upper-majorization-polytope $f$-divergence problems,
respectively, and, through the fidelity reduction, the corresponding
pure-state entanglement and coherence problems. 
}
\label{fig:t3t4}
\end{figure}

Since $m=10<n$, their $m$-dimensional increment vectors must be
extended to $\Real^n$ according to
\begin{equation}
\widevectau^{(3)}
=
\vectau^{(3)}
\oplus
(x_{11},x_{12}),
\qquad
\widevectau^{(4)}
=
\vectau^{(4)}
\oplus
(0,0).
\end{equation}
Figure~\ref{fig:t3t4} shows the corresponding T3 and T4 constructions.

The corresponding optimal representative probability vectors are
\begin{equation}
\begin{aligned}
\vecx_*^{(1)}
&=
\vecy-\vectau^{(1)}
\\
&=
(0.166680,\,0.166680,\,0.128880,\,0.126880,\,
0.126880,\,0.058400,
\\[-1mm]
&\hspace{1.2cm}
0.047200,\,0.047200,\,0.047200,\,0.047200,\,
0.018400,\,0.018400),
\end{aligned}
\end{equation}
and
\begin{equation}
\begin{aligned}
\vecx_*^{(2)}
&=
\vecy-\vectau^{(2)}
\\
&=
(0.272800,\,0.268000,\,0.132350,\,0.130350,\,
0.130350,\,0.027950,
\\[-1mm]
&\hspace{1.2cm}
0.009550,\,0.009550,\,0.009550,\,0.009550,\,
0,\,0).
\end{aligned}
\end{equation}
For T3 and T4, the optimal representatives are the extended increment
vectors themselves:
\begin{equation}
\begin{aligned}
\vecx_*^{(3)}
&=
\widevectau^{(3)}
\\
&\simeq
(0.163368,\,0.163368,\,0.130900,\,0.129182,\,
0.129182,\,0.058400,
\\[-1mm]
&\hspace{1.2cm}
0.047200,\,0.047200,\,0.047200,\,0.047200,\,
0.020000,\,0.016800),
\end{aligned}
\end{equation}
and
\begin{equation}
\begin{aligned}
\vecx_*^{(4)}
&=
\widevectau^{(4)}
\\
&\simeq
(0.272800,\,0.268000,\,0.112947,\,0.111465,\,
0.111465,\,0.035574,
\\[-1mm]
&\hspace{1.2cm}
0.021937,\,0.021937,\,0.021937,\,0.021937,\,
0,\,0).
\end{aligned}
\end{equation}
The entries displayed for T3 and T4 are rounded to six decimal places.

Thus, once the appropriate taut string has been constructed, no
further optimization is required: T1 and T2 yield the optimal
representatives $\vecy-\vectau^{(1)}$ and
$\vecy-\vectau^{(2)}$, whereas T3 and T4 yield
$\widevectau^{(3)}$ and $\widevectau^{(4)}$ directly.
These representative vectors determine the corresponding canonical
optimal states according to Table~II of the Letter.

\medskip
\noindent
\textit{Numerical evaluation for representative measures.---}
We now evaluate the optimal errors for one representative measure from
each class. We choose trace distance as the spectral distance and the
sandwiched R\'enyi divergence of order $\alpha=1/2$ as the quantum
divergence. These choices do not modify the taut strings or their
optimal representatives; they only assign a numerical error to each
solution.

\begin{itemize}[leftmargin=*]

\item \textit{Trace distance.}
For
\begin{equation}
g(\vec v)
=
\frac{1}{2}\sum_i |v_i|,
\end{equation}
the spectral distance $d_g$ is the trace distance. For
non-uniformity, geometries T1 and T2 therefore give
\begin{equation}
\varepsilon_{\L,d_g}^{\min}
=
g\left(\vectau^{(1)}\right),
\qquad
\varepsilon_{\U,d_g}^{\min}
=
g\left(\vectau^{(2)}\right).
\end{equation}

For pure-state entanglement and coherence, the same trace distance
reduces instead to the root-infidelity objective
\begin{equation}
g\!\left(
D_{\RI}(\vecxp,\vecy),
-D_{\RI}(\vecxp,\vecy),
\vec{0}_{D-2}
\right)
=
D_{\RI}(\vecxp,\vecy),
\end{equation}
where
\begin{equation}
D_{\RI}(\vecx,\vecy)
=
\sqrt{1-F^2(\vecx,\vecy)},
\qquad
F(\vecx,\vecy)
=
\sum_i\sqrt{x_i y_i}.
\end{equation}
Because the representation map is order reversing, T4 gives the
optimal representative for the quantum set $\L(\rho)$, while T3 gives
the optimal representative for $\U(\rho)$.

\item \textit{Sandwiched R\'enyi divergence of order $\alpha=1/2$.}
For non-uniformity, we choose the sandwiched R\'enyi divergence
$\widetilde D_{1/2}$, whose unitary-orbit reduction is
\begin{equation}
\widetilde D_{1/2}(\vecx\|\vecy)
=
-2\log F(\vecx,\vecy).
\end{equation}
For the convex generator
\begin{equation}
f_{\mathrm H}(t)
=
(\sqrt{t}-1)^2,
\end{equation}
the corresponding Csisz\'ar $f$-divergence satisfies
\begin{equation}
D_{f_{\mathrm H}}(\vecx\|\vecy)
=
2\bigl[1-F(\vecx,\vecy)\bigr],
\end{equation}
and therefore
\begin{equation}
\widetilde D_{1/2}(\vecx\|\vecy)
=
h\!\left(
D_{f_{\mathrm H}}(\vecx\|\vecy)
\right),
\qquad
h(u)
=
-2\log(1-u/2).
\end{equation}
Since $h$ is strictly increasing, this evaluates precisely the same T3 and T4 optimal representatives constructed above. 
Notice that $f_{\mathrm H}$ is also the generator used in the proof of main theorem in the Letter to connect the T3/T4
construction with the root-infidelity objective.

Unlike quantum relative entropy, $\widetilde D_{1/2}$ remains finite
for the present non-full-support target distribution $\vecy$ whenever
the corresponding fidelity is nonzero, as is the case here.

\end{itemize}

The resulting optimal errors are collected in
Table~\ref{tab:results}.

\begin{table}[!htbp]
\centering
\begin{tabular}{lcccc}
\toprule
MQRT & Measure & Set & Geometry & $\varepsilon^{\min}$ \\
\midrule
\multirow{4}{*}{Non-uniformity}
& \multirow{2}{*}{$d_g$ (trace distance)}
& $\L(\rho)$ & T1 & $0.1176$ \\
&
& $\U(\rho)$ & T2 & $0.1604$ \\
\cmidrule(lr){2-5}
&
\multirow{2}{*}{$S=\widetilde D_{1/2}$}
& $\L(\rho)$ & T3 & $0.0506$ \\
&
& $\U(\rho)$ & T4 & $0.0262$ \\
\midrule
\multirow{2}{*}{Entanglement or coherence}
& \multirow{2}{*}{$d_g$ (trace distance)}
& $\L(\rho)$ & T4 & $0.1609$ \\
&
& $\U(\rho)$ & T3 & $0.2221$ \\
\bottomrule
\end{tabular}

\caption{
Optimal approximate-conversion errors obtained from the four
taut-string solutions of Figs.~\ref{fig:t1t2} and
\ref{fig:t3t4}. Trace distance is used as a representative spectral
distance and $\widetilde D_{1/2}$ as a representative quantum
divergence. The ``Set'' column refers to the quantum conversion set.
For pure-state entanglement and coherence, the order-reversing
representation map associates $\L(\rho)$ with the classical polytope
$\U(\vecx)$ and $\U(\rho)$ with $\L(\vecx)$, hence the appearance of
T4 and T3, respectively. For T3 and T4, the reported values are
evaluated on the extended representatives
$\widevectau^{(3)}$ and $\widevectau^{(4)}$.
}
\label{tab:results}
\end{table}

As a consistency check, for the two T3/T4 solutions the representative
measures used above satisfy
\begin{equation}
\widetilde D_{1/2}
=
-\log\!\left(1-D_{\RI}^2\right).
\end{equation}
Thus, the T4 values
$D_{\RI}\simeq0.1609$ and
$\widetilde D_{1/2}\simeq0.0262$, and the T3 values
$D_{\RI}\simeq0.2221$ and
$\widetilde D_{1/2}\simeq0.0506$, correspond to the same optimal
representative vectors. This illustrates explicitly that the
distinguishability measure affects only the final numerical value of
the error, while the underlying taut-string geometry remains
unchanged.

\bibliography{references}